\documentclass[structabstract]{aa}  
\usepackage{url}
\usepackage{natbib}
\usepackage{graphicx}
\usepackage{txfonts}

\newcommand{\Mp}{\ensuremath{M_\mathrm{p}}}
\newcommand{\Ms}{\ensuremath{M_\mathrm{s}}}

\newcommand{\Kp}{\ensuremath{K_\mathrm{a}}}
\newcommand{\Ks}{\ensuremath{K_\mathrm{b}}}
\newcommand{\Aap}{\ensuremath{a_\mathrm{app}}}
\newcommand{\Aph}{\ensuremath{a_\mathrm{phy}}}

\newcommand{\Msun}{\ensuremath{\mathrm{M_{\sun}}}}

\newcommand{\excs}{\extracolsep{\fill}}

\newcommand{\dg}{$^\circ$} 
\newcommand{\so}{$_{\odot}$} 
\newcommand{\hd}{HD~200775}

\newcommand{\Ha}{H$_{\alpha}$~}

\begin{document} %
%   \title{Origin of the H$_{\alpha}$ line in the young massive binary HD~200775 }% \thanks{Based on observations made with
 %  \title{Enhanced ejection in the young spectroscopic binary HD~200775 \thanks{Based on observations made with the VEGA/CHARA instrument}}
   \title{Enhanced \Ha activity at periastron in the young and massive spectroscopic binary HD~200775 \thanks{Based on observations made with the VEGA/CHARA instrument}}

   % \subtitle{xxx}
\author{M.~Benisty\inst{1,2}, K.~Perraut\inst{1}, D.~Mourard\inst{3}, P.~Stee\inst{3}, G.H.R.A.~Lima\inst{1}, J.B.~Le~Bouquin\inst{1}, M.~Borges~Fernandes\inst{4},  O.~Chesneau\inst{3}, N.~Nardetto\inst{3}, I.~Tallon-Bosc\inst{5}, H.~McAlister\inst{6,7}, T.~Ten~Brummelaar\inst{7}, S.~Ridgway\inst{8}, J.~Sturmann\inst{7}, L.~Sturmann\inst{7}, N.~Turner\inst{7}, C.~Farrington\inst{7}, P.J.~Goldfinger\inst{7}}

\institute{        Institut d'Astrophysique et de Plan\'etologie de Grenoble, CNRS-UJF UMR 5571,
          414 rue de la Piscine, 38400 St Martin d'H\`eres, France 
          \and
           Max Planck Institut f\"ur Astronomie, Konigst\"uhl 17, 69117
          	Heidelberg, Germany 
           \and
	Laboratoire Lagrange, UMR 7293 UNS-CNRS-OCA, Boulevard de l'Observatoire, B.P. 4229 F, 06304 NICE Cedex 4, France
	\and
          Observatorio Nacional, Rua General Jos\'e Cristino, 77, 20921-400, Sao Cristovao, Rio de Janeiro, Brazil
          \and
          Universit\'e de Lyon, 69003 Lyon, France; Universit\'e Lyon 1, Observatoire de Lyon, 9 avenue Charles Andr\'e, 69230 Saint Genis
Laval; CNRS, UMR 5574, Centre de Recherche Astrophysique de Lyon; Ecole Normale Sup\'erieure, 69007 Lyon, France
          \and
          Georgia State University, PO Box 3969, Atlanta GA 30302-3969, USA 
          \and
          CHARA Array, Mount Wilson Observatory, 91023 Mount Wilson CA, USA
         \and
   	National Optical Astronomy Observatory, PO Box 26732, Tucson, AZ 85726, USA\\}        

         \offprints{Myriam.Benisty@obs.ujf-grenoble.fr} 
         \date{Received 26 June 2012 / Accepted 7 May 2013}

  \abstract
 {Young close binaries clear central cavities in their surrounding circumbinary disk from which the stars can still accrete material. This process takes place within the very first astronomical units, and is still not well constrained as the observational evidence has been gathered, until now, only by means of spectroscopy. }
   % Herbig Be stars are intermediate-mass young stars surrounded by a complex circumstellar environment. Like their lower mass counterparts, the T Tauri stars, a large fraction lies in multiple systems. \textbf{Mechanisms formation \& timescale involved in formation of SB. Enjeux? Questions?} Studying multiple systems is therefore of strong interest to better understand the formation processes, the mechanisms that affect their circumstellar environment and constrain the stellar evolution. }
   {The young object HD~200775 (MWC~361) is a massive spectroscopic binary (separation of $\sim$15.9~mas, $\sim$5.0~AU), with uncertain classification (early/late Be), that shows a strong and variable \Ha emission. We aim to study the mechanisms that produce the \Ha line at the AU-scale. }
    {Combining the  radial velocity measurements and astrometric data available in the literature, we determined new orbital parameters. With the VEGA instrument on the CHARA array, we spatially and spectrally resolved the \Ha emission of \hd~on a scale of a few milliarcseconds,  at low and medium spectral resolutions (R$\sim$1600 and 5000) over a full orbital period ($\sim$3.6 years). }
   {We observe that the \Ha equivalent width varies with the orbital phase, and increases close to periastron, as expected from theoretical models that predict an increase of the mass transfer from the circumbinary disk to the primary disk. In addition, using spectral visibilities and differential phases, we find marginal variations of the typical extent of the \Ha emission (at 1 to 2$\sigma$ level) and location (at 1 to 5$\sigma$ level). The spatial extent of the \Ha emission, as probed by a Gaussian FWHM, is minimum at the ascending node (0.67$\pm$0.20~mas, i.e., 0.22$\pm$0.06~AU), and more than doubles at periastron. In addition, the Gaussian photocenter is slightly displaced in the direction opposite to the secondary, ruling out the scenario in which all or most of the \Ha emission is due to accretion onto the secondary. These findings, together with the wide \Ha line profile, may be due to a non-spherical wind enhanced at periastron.}
   {For the first time in a system of this kind, we spatially resolve the \Ha line and estimate that it is emitted in a region larger than the one usually inferred in accretion processes. The \Ha line could be emitted in a stellar or disk-wind, enhanced at periastron as a result of gravitational perturbation, after a period of increased mass accretion rate. Our results suggest a strong connection between accretion and ejection in these massive objects, consistent with the predictions for lower-mass close binaries.} %Additional datasets are however needed to disentangle the exact process of mass ejection (disk versus stellar wind) and ensure a better characterization of the stellar properties. }

   \keywords{Methods: observational Techniques: high angular
   resolution - Techniques: interferometric - Stars: binary (HD~200775)
   Stars: emission-line - Stars: circumstellar matter}

    \authorrunning {M.~Benisty et al.}
   \titlerunning{Enhanced \Ha activity at periastron in the young and massive spectroscopic binary HD~200775}
   \maketitle
%________________________________________________________________

\section{Introduction}

Herbig AeBe (HAeBe) stars are pre-main-sequence objects of intermediate mass, with spectral types from B to F. 
They are surrounded by protoplanetary disks of gas and dust, responsible for the observed excess
emission from the infrared to the submillimeter \citep[e.g.,][]{alonso09}. Their spectra display many emission lines, which are signatures of accretion and outflow \citep{mundt94}. These objects are of particular interest as they lie between solar-mass young stars, thought to form through gravitational collapse of a molecular cloud, and the massive young stars for which the process of formation is still matter of debate. 

Like their lower mass counterparts \citep[e.g.,][]{duchene07}, a large fraction of Herbig AeBe stars is found in multiple systems (up to 68$\pm$11\%, \citet{baines06}).  
Once formed, close binary stars are thought to accrete mass from an envelope through a circumbinary disk. However, it is not clear what effect preferential mass accretion  will have on the evolution of the two components. Observations have shown that a number of close T~Tauri star binaries \citep[e.g., DQ Tau,][]{basri97} show enhanced emission line activity close to periastron, indicating that the accretion is non-axisymmetric.  Numerical studies of young close binary systems have shown that an inner cavity forms inside the 2:1 resonance and that accretion streamers can still feed the stars inside the circumbinary disk, producing the observed periodic line changes  \citep{artymowicz96,gunther02,devalborro11}.  This interaction between the circumbinary disk and the stars occurs at (sub-)AU scales and, until now, only spatially unresolved observations have been published.  Being able to directly probe the disk-binary interaction is crucial to test models of young binary evolution.

The object \hd~(MWC361) is a triple system consisting of a spectroscopic binary (SB) at $\sim$18~milliseconds of arc (mas) separation \citep{millangabet01}, and a third companion at 6" \citep{li94}.  Based on the analysis of the H$_{\alpha}$ line,  a radial velocity (RV) orbit with a period of 1341 days was reported for the SB \citep{pogodin04}. From their analysis of 33 spectral features, \citet{hernandez04} classified the SB as a Herbig~Be star with a luminosity of 15000-L\so.  However, because of the large uncertainty on the age of the system, the fundamental parameters of the individual sources remain highly uncertain, although it is likely that at least one of the two is an early Herbig~Be~star that dominates the spectrum.  

Various terminologies and criteria have been used to distinguish one star from the other in the literature, generating some confusion.
Using H-band interferometric measurements, \citet{monnier06} determined an astrometric orbit with a projected separation of
15.14~$\pm$~0.70~mas and an inclination of i$\sim$65\dg$\pm$8\dg. They defined as primary target, the brightest near infrared (NIR) source, and found an H-band brightness ratio of 6.5~$\pm$~0.5. They modeled the NIR visibilities around the primary star, assuming that the secondary was unresolved in the NIR, i.e., it did not possess any extended disk at their angular resolution ($\sim$4.3~mas). Using analytical models, they found a uniform disk diameter of 3.6~$\pm$~0.5~mas, i.e., 1.3~$\pm$~0.2~AU, at a  distance of 360$^{+120}_{-70}$ pc. 

\hd~was later observed during an extensive spectropolarimetric campaign  \citep{alecian08}. Two individual components were found in the photospheric line profiles. The authors defined as the primary target the one that emits the sharper lines: the secondary was determined to be the one responsible for the broader, shallower lines. Based on the RV of the lines, the authors provided orbital parameters in overall agreement with \citet{pogodin04} and \citet{monnier06}, and a mass ratio primary/secondary of 0.81$\pm$0.22, indicating that the star considered as secondary is in practice the most massive one. The \Ha bisector velocities computed at a line intensity of 1.5 times the continuum were found to trace the RV of the secondary, suggesting that the line emission is dominated by this star. Their observations revealed a strongly inclined dipolar magnetic field (1000~$\pm$~150~G), found to have been stable for more than 2 years, and related to the object considered as the primary. Although the authors derive similar masses and effective temperatures ($\sim$10~M$_{\odot}$ and 18600~K, respectively), the discrepancy between the observational properties suggests that the two stellar components must have grown and evolved differently.  %secondary is more extinct than the primary. %\citet{alecian08} suspect that % , and derive luminosity estimates for each source using the Stokes I profiles and the total luminosity previously estimated \citep{hernandez04}. %From the luminosities, and evolutionary tracks, they derive mass estimates of 10.7$\pm$2.5~M\so~and 9.3$\pm$2.1~M\so~for the primary and secondary, respectively. The authors claim that the discrepancy between the consequent mass ratio and the one derived using the RV may be due to the inaccurate estimate of the individual luminosities. %The distance of the system is in addition poorly constrained. Various estimates can be found in the literature. 

It seems very likely that the component defined as secondary in \citet{alecian08} is actually the one defined as primary in \citet{monnier06}, the most massive component. 
The RV mass estimate and the dominant \Ha activity is consistent with the presence of a circumstellar dusty disk as invoked in  \citet{monnier06} around the most massive component.  In this paper, we consider the primary to be the most massive component, i.e., the one that possesses a circumstellar dusty disk and dominates the \Ha emission. The secondary is the least massive component that possesses a strong magnetic field. 

%\textbf{In addition, we use the orbital parameters and dynamical distance derived in this work by the combination of both astrometric measurements \citep{monnier06} and radial velocity data \citep{alecian08}.}
 
Mid-infrared (MIR) images obtained with the Keck segment-tilting experiment indicated a large halo containing 45\% of 10.7~$\mu$m flux with a north-south elongation \citep{monnier09} consistent with the orientation of the binary orbit measured by \citet{monnier06}. This suggests that the halo is the remnant of a circumbinary disk. This was confirmed by Subaru MIR images that showed a diffuse emission with an elliptical shape, suggesting an inclined flared disk \citep[i=54.5\dg$\pm$1.2\dg;][]{okamoto09}. The MIR emission extends up to 20 times the semi-major axis of the binary, which indicates a large gap in the system. Furthermore, the system lies in a large scale biconical cavity that has very likely been excavated by an extended bipolar outflow inclined by $\sim$70$^\circ$ \citep{fuente98, watt86}, a value also close to the orbital plane inclination. These results support the presence of a circumbinary  disk in the same plane as the orbit.  % The authors concluded that the MIR emission likely trace the surface of a photo-evaporation wind.

The H${\alpha}$ and H$\beta$ emission lines show great variations over time, indicating changes in activity with a period of 3.68
years, in agreement with the binary period \citep{pogodin04}. In low states, the lines are double-peaked. In active states,
the line intensities and equivalent widths (EW) increase, while their profiles show a complicated multi-component structure, including
a doubling of the central absorption feature with a new, variable, blue-shifted component in addition to the pre-existing red-shifted one. The EW is found to be at its maximum right after periastron, indicating that the line activity is indeed related to the binarity.

The hot gas, responsible for the Hydrogen line, can be involved in accretion and ejection flows close to the source and can be used to probe the corresponding physical conditions. These phenomena, however, occur in a small region of a few AU around the star, corresponding to a few mas.  With the recent advent of spectro-interferometric instruments, it has been possible to achieve such a resolution, and to spatially and spectrally resolve some of these lines.  The first studies of the kind showed that the Br$_{\gamma}$ emission line was probably tracing winds or gas in a rotating disk  \citep{malbet07,kraus08}. One of the challenging goals is to study the launching points of the jets and discriminate between the various theoretical models, X-wind \citep{shu94} and disk-wind \citep{casse00}. Inspired by the magneto-centrifugal scenario for the acceleration of jets, \citet{perraut10} and \citet{weigelt11} provided realistic solutions to account for the line emission in disk winds. 

From 2008 to 2011, we led an observing campaign over an entire orbit of \hd~with the optical spectro-interferometer VEGA installed at the CHARA Array. The paper is organized as follows: in Sect.~\ref{sec:obs} we describe the observations and the data processing. We present the new orbital solution, spectra, and interferometric observables in Sects.~\ref{sec:orb} and \ref{sec:interfero}, and describe our modeling in Sect.~\ref{sec:res}. We discuss our results in Sect.~\ref{sec:disc} and conclude in Sect.~\ref{sec:conc}.

\begin{table*}[t] \centering \caption{Log of the observations.  $\phi$ is the orbital phase (zero at periastron). Italics indicate that only spectra were retrieved.} \label{tab:log}
\begin{tabular}{ccccccccccccc}
\hline
Date & UT & HJD & HA & Bp & PA$_{\rm{B}}$ & Spectral & Calibrator & $r_0$ &  $\phi$\\%& $\Phi$ & $\Phi$ \\
 & (h) & (2 450 000+)& (h) & (m) & ($^\circ$) & resolution & (HD number) & (cm)  &  \\%& (\dg) & (mas)\\
\hline
\hline
2008-07-29 & 10.26 & 4676.9271  & 1.86 & 26.7 & -31.0 & 1600 & 204770 & 8-10 &  -0.012 \\%&  -10 & 0.14\\
2008-08-07 & 9.41 & 4685.8924  & 1.60 & 27.0 & -28.0 & 1600 & 204770 & 7-8 &  -0.006 \\%& -15  & 0.21\\
\textit{2008-11-24} & \textit{4.13} & \textit{4794.6722}  & \textit{--} & \textit{--} & \textit{--} & \textit{5000} & \textit{--} & 5 &  \textit{0.070}\\%& -- & --\\
\textit{2009-07-26} & \textit{11.86} & \textit{5038.9944} & \textit{--} & \textit{--} & \textit{--} & \textit{1600} & \textit{--} & 5-6 & \textit{0.240}\\%& -- & --\\
\textit{2009-08-02} & \textit{8.42} & \textit{5045.8507}  & \textit{--} & \textit{--} & \textit{--} & \textit{1600} & \textit{--} &5-6 &  \textit{0.245}\\%& -- & --\\
2009-08-26 & 8.46 & 5069.8228 &  1.86 & 26.8 & -31.0 & 1600 & 197950 & 8 &0.262\\%   & 0 & 0\\
2010-08-01 & 7.00 & 5409.7917 & -1.25 & 28.3 & 0.9 & 5000 & 197950 ; 204770 & 10-11 &  0.499\\% & 15  & 0.22 \\
\textit{2010-09-14} & \textit{6.95} & \textit{5453.7833} & \textit{--} & \textit{--} & \textit{--} & \textit{5000} & \textit{--} & 7 &  \textit{0.530}\\% & -- & --\\
\textit{2010-09-17} & \textit{8.07} & \textit{5456.8361}  & \textit{--} & \textit{--} & \textit{--} & \textit{5000} & \textit{--} & 6-7 & \textit{0.532}\\%& -- & --\\
\textit{2011-07-28} & \textit{9.61} & \textit{5770.8896} & \textit{--} & \textit{--} & \textit{--} & \textit{5000} & \textit{--} & 5 &  \textit{0.752}\\%& -- & --\\
\textit{2011-07-29} & \textit{7.77} & \textit{5771.8236}  & \textit{--} & \textit{--} & \textit{--} & \textit{5000} & \textit{--} & 7 &  \textit{0.752}\\%& -- & --\\
2011-08-28 & 9.63 & 5801.9007 & 3.11 & 25.1 & -44.0 & 5000 & 197950 & 6-7 &  0.773\\%& 0 & 0\\
2011-10-16 & 5.61 & 5850.7340 & 2.31 & 26.2 & -36.0 & 5000 & 197950 & 7-8 &  0.806\\%& 20 & 0.29\\
\textit{2011-10-21} & \textit{3.62} & \textit{5855.6507} & \textit{--} & \textit{--} & \textit{--} & \textit{5000} &\textit{--}& 5-6 & \textit{0.811} \\% & -- & --\\
\hline
\end{tabular}
\end{table*}

\section{Observations and data processing} \label{sec:obs}

\subsection{VEGA observations}
The data were collected at the CHARA array \citep{chara}, with the VEGA instrument \citep{vega}. Our datasets were obtained with two telescopes (S1S2) and cover the period from July 2008 to October 2011, i.e., an orbital period.  
The average projected baseline length (Bp) is about 27~m, and the average baseline position angles (PA$_{\rm{B}}$) are given in Table~\ref{tab:log}. The angular resolution ($\lambda$/2Bp) of our observations is $\sim$2.5 mas. We therefore fully resolve the spectroscopic binary. The interferometric field of view is $\pm$2" in the slit direction and excludes the third object of the system (at 6").

The first dataset was recorded with the lowest spectral resolution of VEGA (R$\sim$1600, hereafter LR), as the target is at the instrumental sensitivity limit ($m_V$~=~7.4). Instrumental  improvements allowed us to later record data in medium spectral resolution (R$\sim$5000, hereafter MR). 

Each observation followed a calibrator-star-calibrator sequence, with 40 files of 1000 short exposures (15 ms) per observation. The calibrators were chosen to be close to the target both in distance and in spectral type, to be small enough at visible wavelengths, and to have an angular diameter known to an accuracy of a few percentage points. Using the SearchCal JMMC tool, we selected HD~204770 and HD~197950 (uniform disk diameters of $\sim$0.17$\pm$0.01~mas and $\sim$0.33$\pm$0.02~mas, respectively).

\subsection{Data processing}
\label{sect:dataproc}

\textbf{\textit{Spectra:}} the spectra were extracted using a classical scheme of collapsing the 2D flux in one spectrum, calibrating the pixel-wavelength relation using a Thorium-Argon lamp, and normalizing the continuum by a polynomial fit. We used the H$_\alpha$ absorption lines of the calibrators to check the spectral calibration at medium resolution. The accuracy of the spectral calibration is 0.13 nm (i.e., 60 km/s)  in MR, and 0.39 nm (i.e., 178 km/s) in LR.\\

\noindent \textbf{\textit{Visibilities and differential phases:}}
The standard routines of the VEGA data reduction pipeline were used \citep[][]{vega}. We computed visibilities and differential phases in individual spectral channels, using the cross-spectrum method between two spectral channels [1] and [2] \citep[for more details, see][]{berio99}. To reach a sufficient signal-to-noise ratio (SNR) (at least 1 photon per speckle, spectral channel and single exposure), we considered the spectral band [1] to be as wide as the entire spectral range (32~nm, i.e., a broad band measurement), and [2], to be 4~nm-wide. 
This method first led  to the determination of the differential phase between [1] and [2], and of the product of the visibility amplitude V$_1$*V$_2$. The dataset was calibrated from the residual atmospheric piston and chromatic optical path difference with the model described in \cite{vega}. ${V_{1}}^2$ were estimated using the integration of the spectral densities of the short exposures over the entire spectral range [1] and calibrated from the instrumental transfer function estimated on the calibrators. Using an estimate of V$_{1}$ and of the transfer function, we deduced the calibrated visibility V$_{2}$ for the narrow spectral channel [2]. By sliding the narrow spectral channel [2] with steps of 2~nm across the entire spectral range [1], we obtained a set of 13 visibilities V$_{2}$ and differential phases, $\Delta\phi =\phi_{2}- \phi_{1}$, with a final spectral resolution of $\sim$160. We considered $\Delta\phi$ equal to 0 in the continuum part of the spectrum, which means that the differential phases measured in the \Ha line correspond to astrometric offsets along the baseline direction, with the photocenter in the continuum as a reference point.  

The data reduction pipeline provided individual errors for each spectral channel, that account for photon noise only. To account for other sources of noise, we adopted conservative errors by considering the largest value between the rms in the continuum and the error  computed by the pipeline.  As the number of photons $N_{H\alpha}$ is much higher in the H$\alpha$ spectral channel than in the continuum, we estimated the error  in the H$\alpha$ visibility by dividing the continuum error  by $\sqrt{N_{H\alpha}/N_{\rm{cont}}}$.  Finally, as HD~200775 is a faint target compared to the VEGA limiting magnitude, excellent weather conditions are required to obtain good quality data. As a consequence, only 6 observations over our 14 attempts have led to a sufficient SNR to produce  interferometric observables.

\begin{figure*}[t] \centering
\begin{tabular}{ccc}
\includegraphics[width=0.4\textwidth]{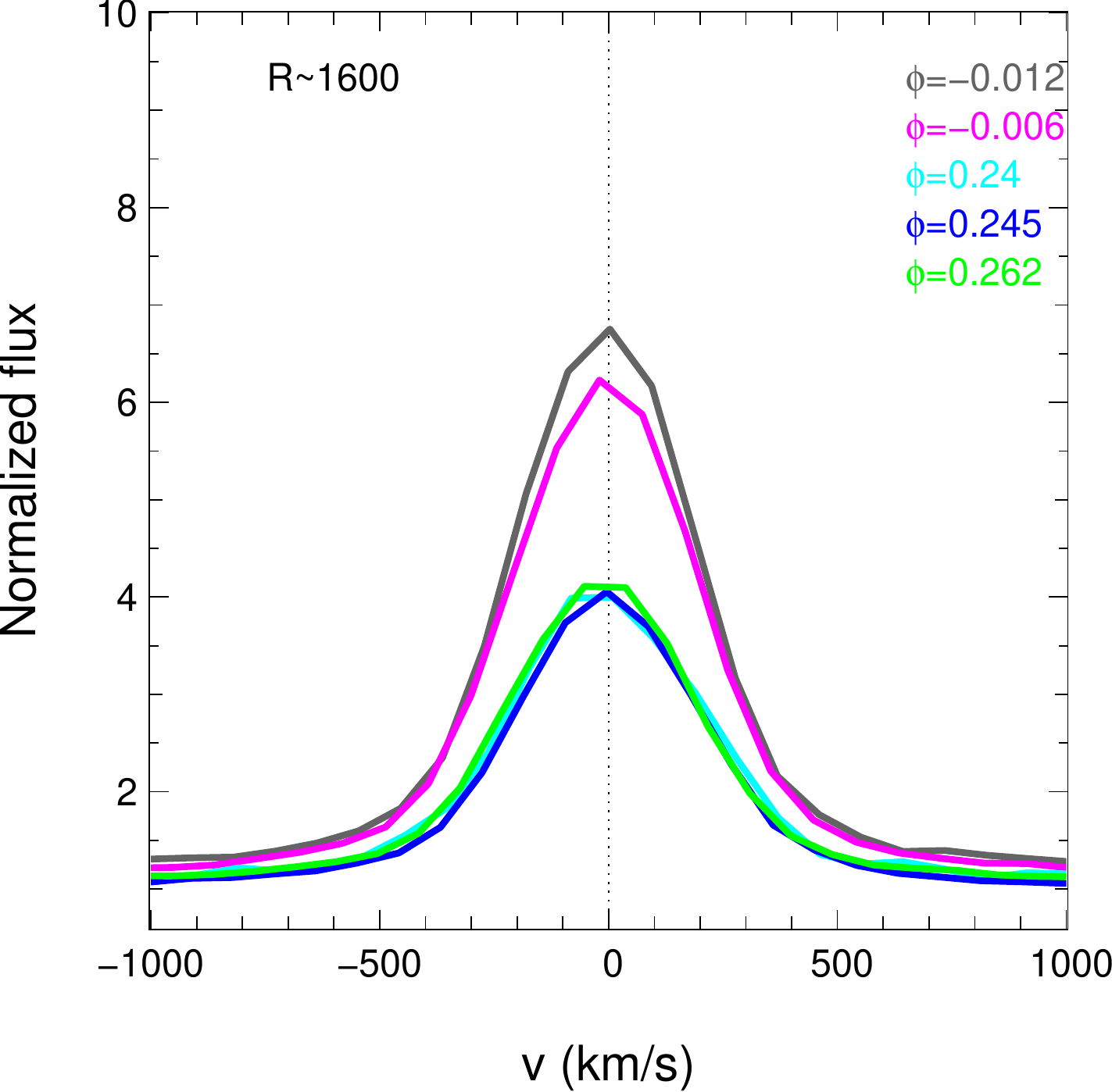}&
\includegraphics[width=0.4\textwidth]{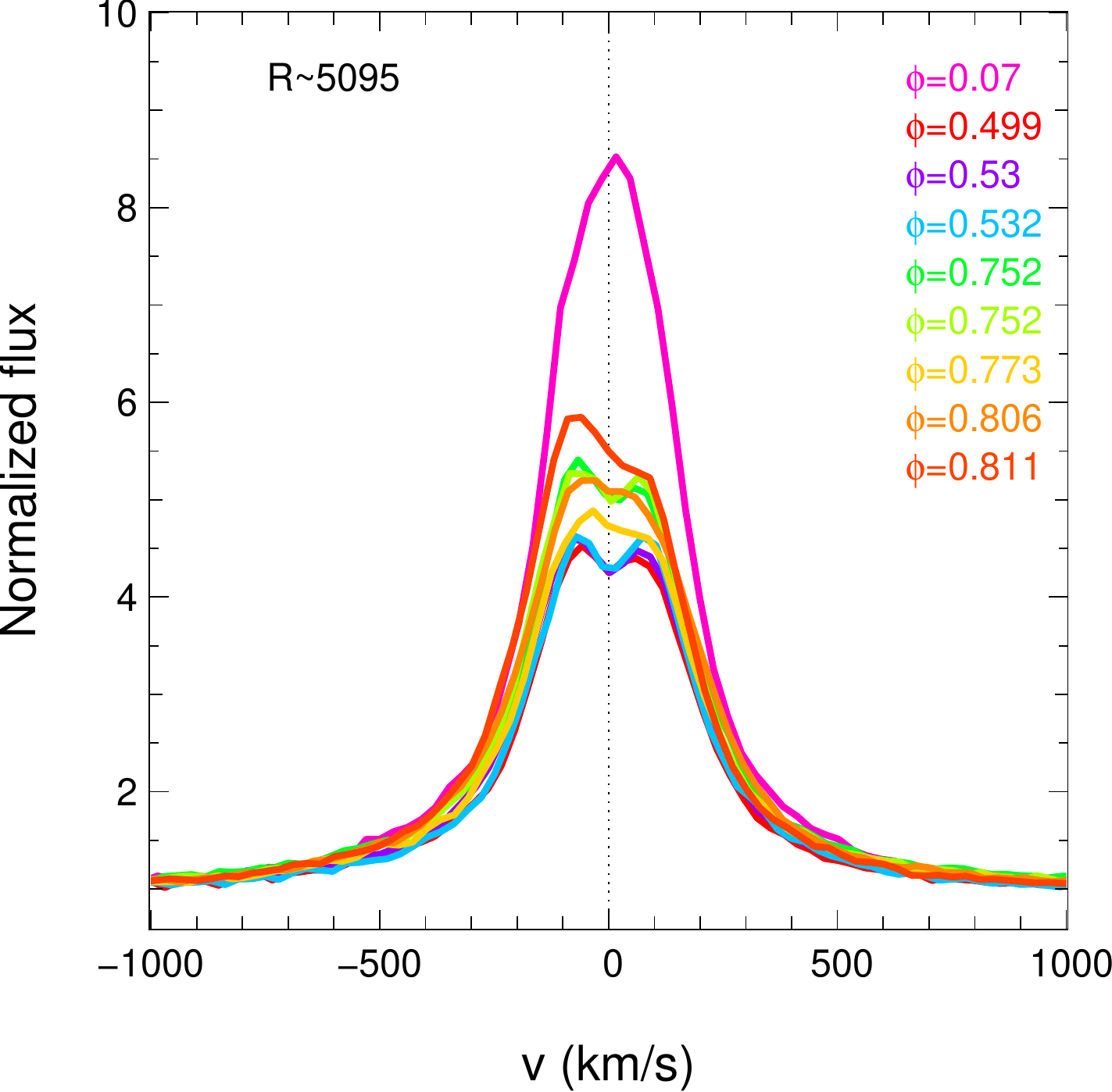}  
%& \includegraphics[width=0.32\textwidth]{EW_Phase_3}
\end{tabular}
\caption{Left:  H$_{\alpha}$ VEGA spectra obtained at R=1600 (LR).  Right: Additional VEGA spectra obtained at R=5000 (MR).}
  \label{fig:spectre}
\end{figure*}

\begin{figure}[t] \centering
\includegraphics[width=0.4\textwidth]{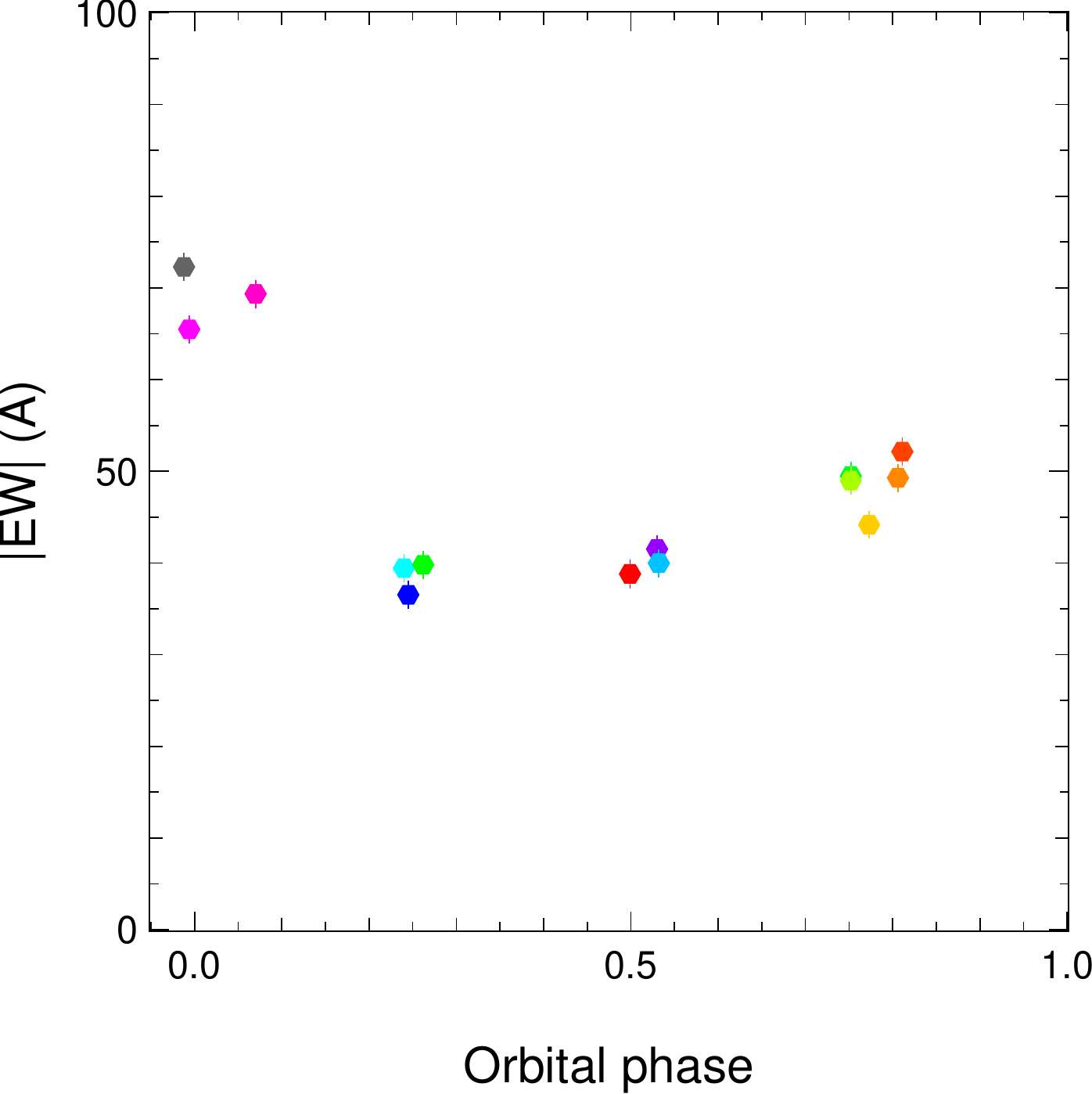}
\caption{Equivalent widths of the H$\alpha$ line over the orbit for all the spectra.}
  \label{fig:EW}
\end{figure}

\begin{figure*}[t] \centering
\begin{tabular}{ccc}
\includegraphics[width=0.32\textwidth]{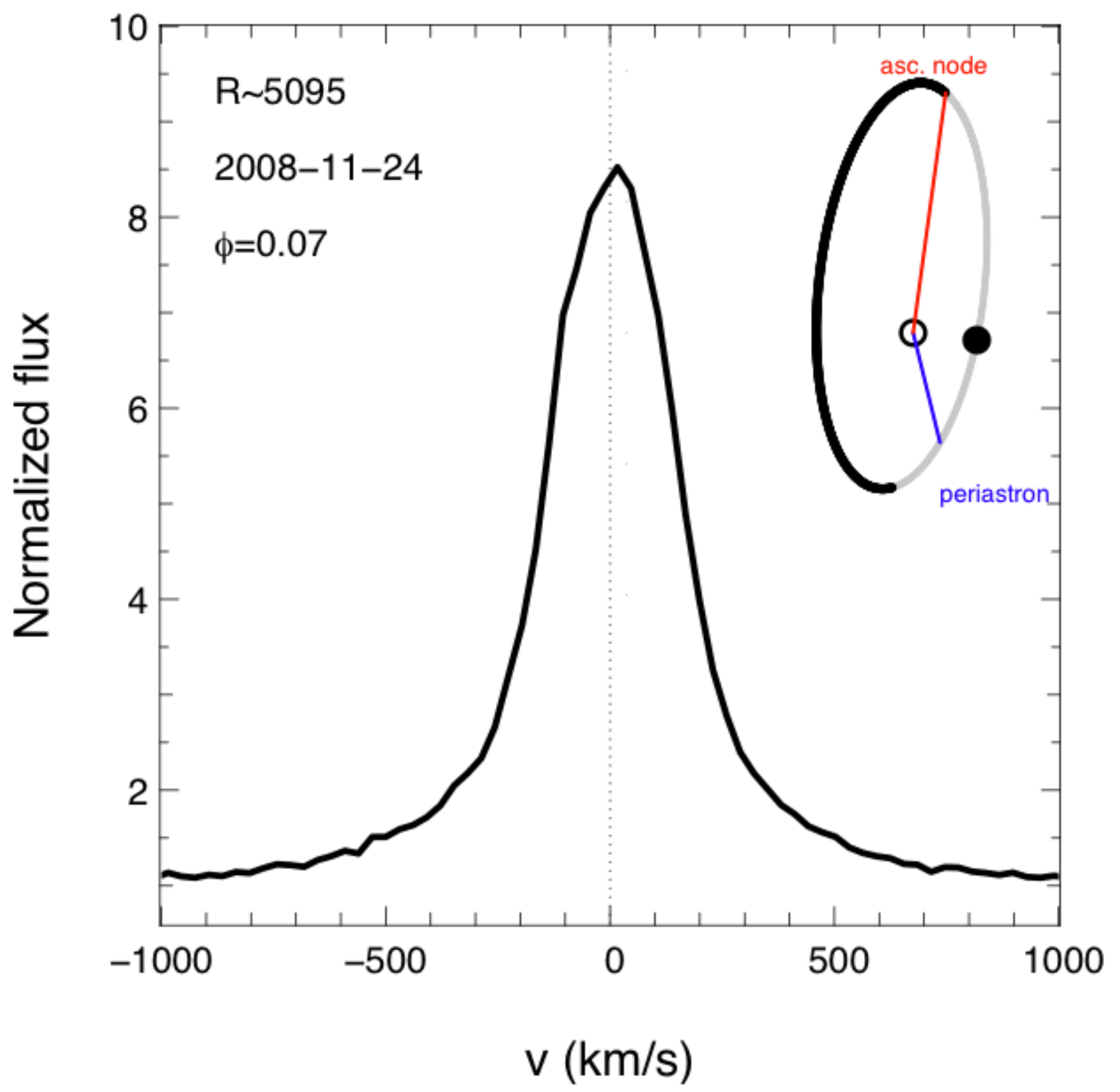} & 
\includegraphics[width=0.32\textwidth]{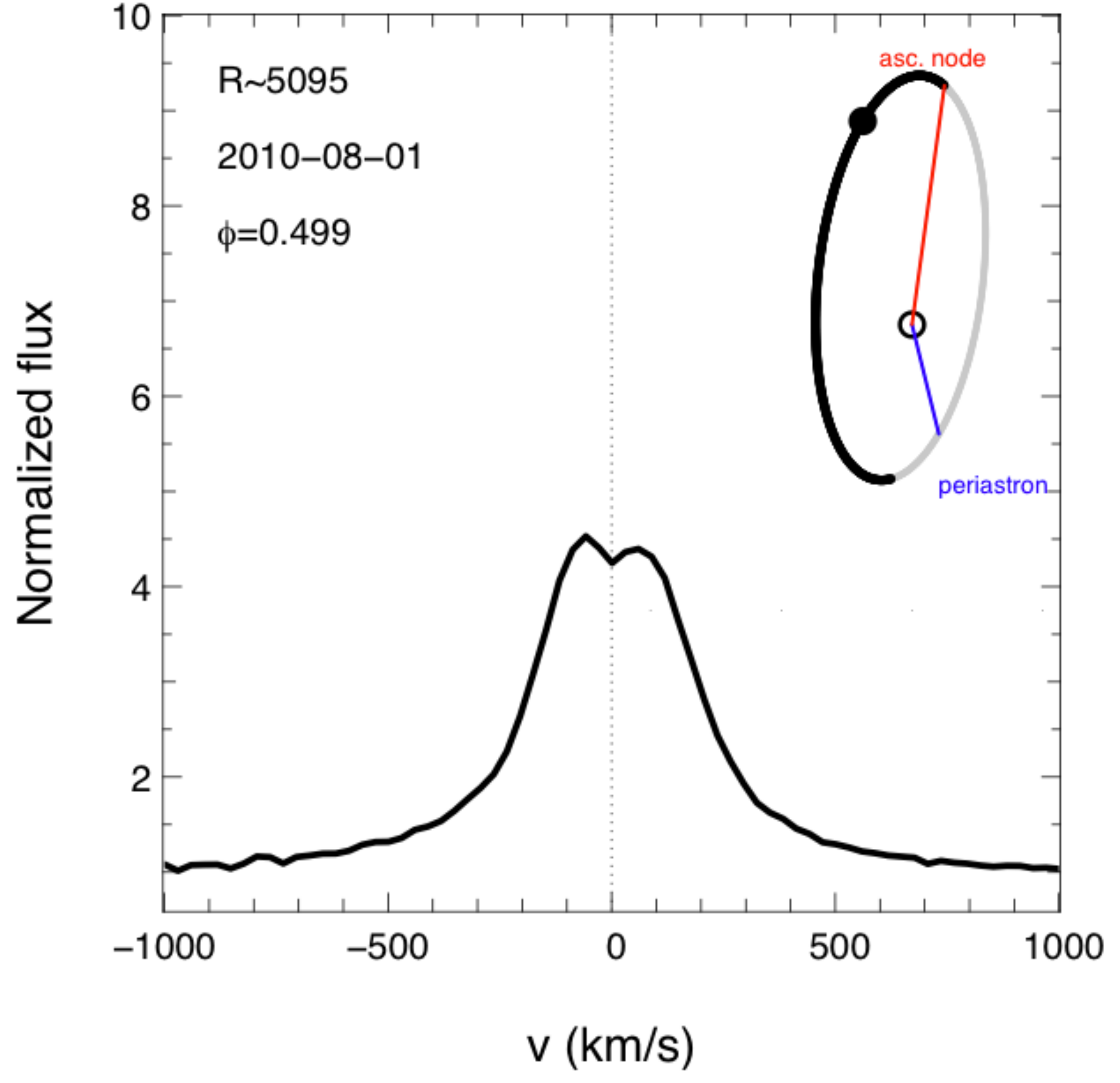}&
\includegraphics[width=0.322\textwidth]{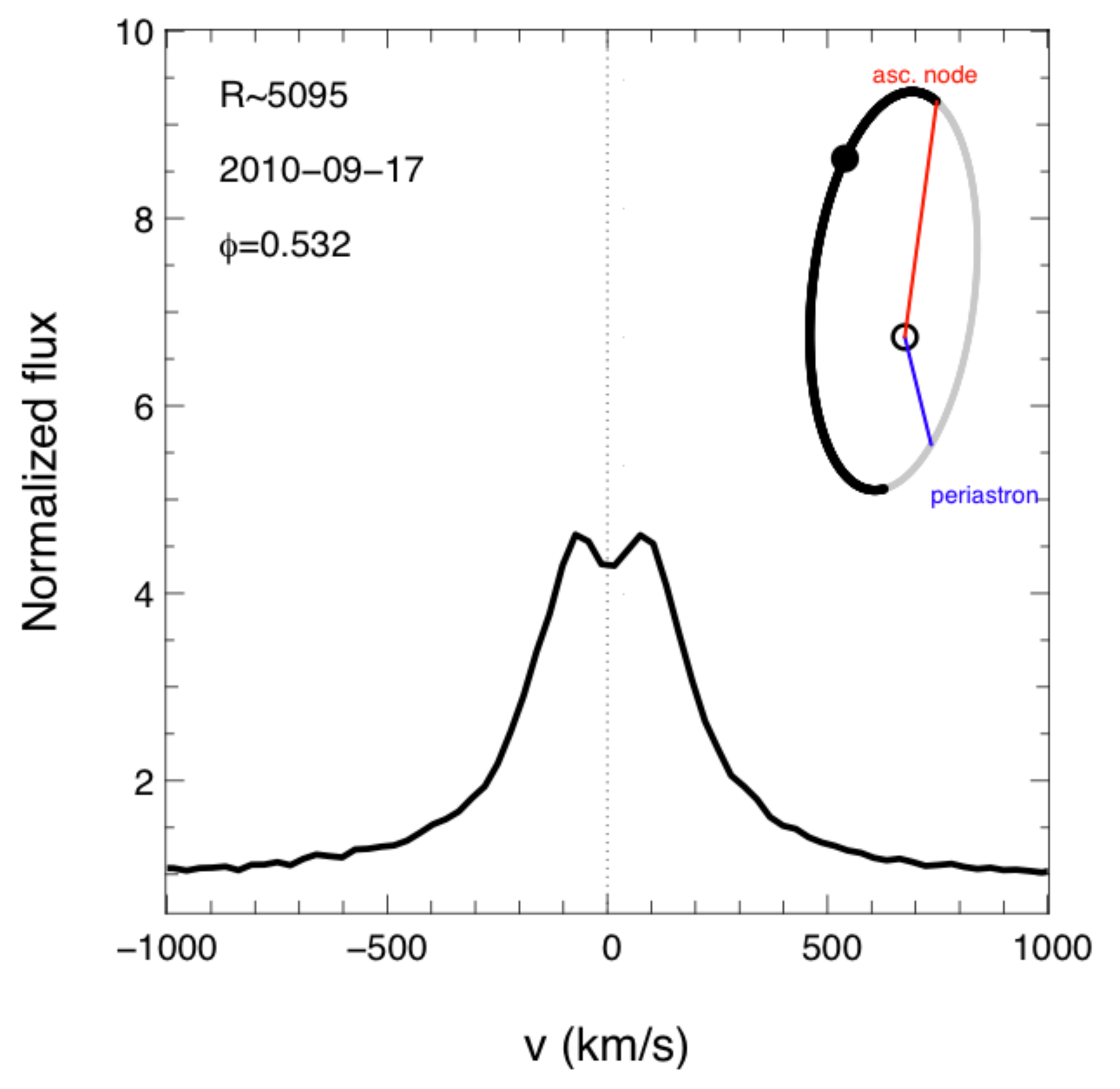}\\
\includegraphics[width=0.323\textwidth]{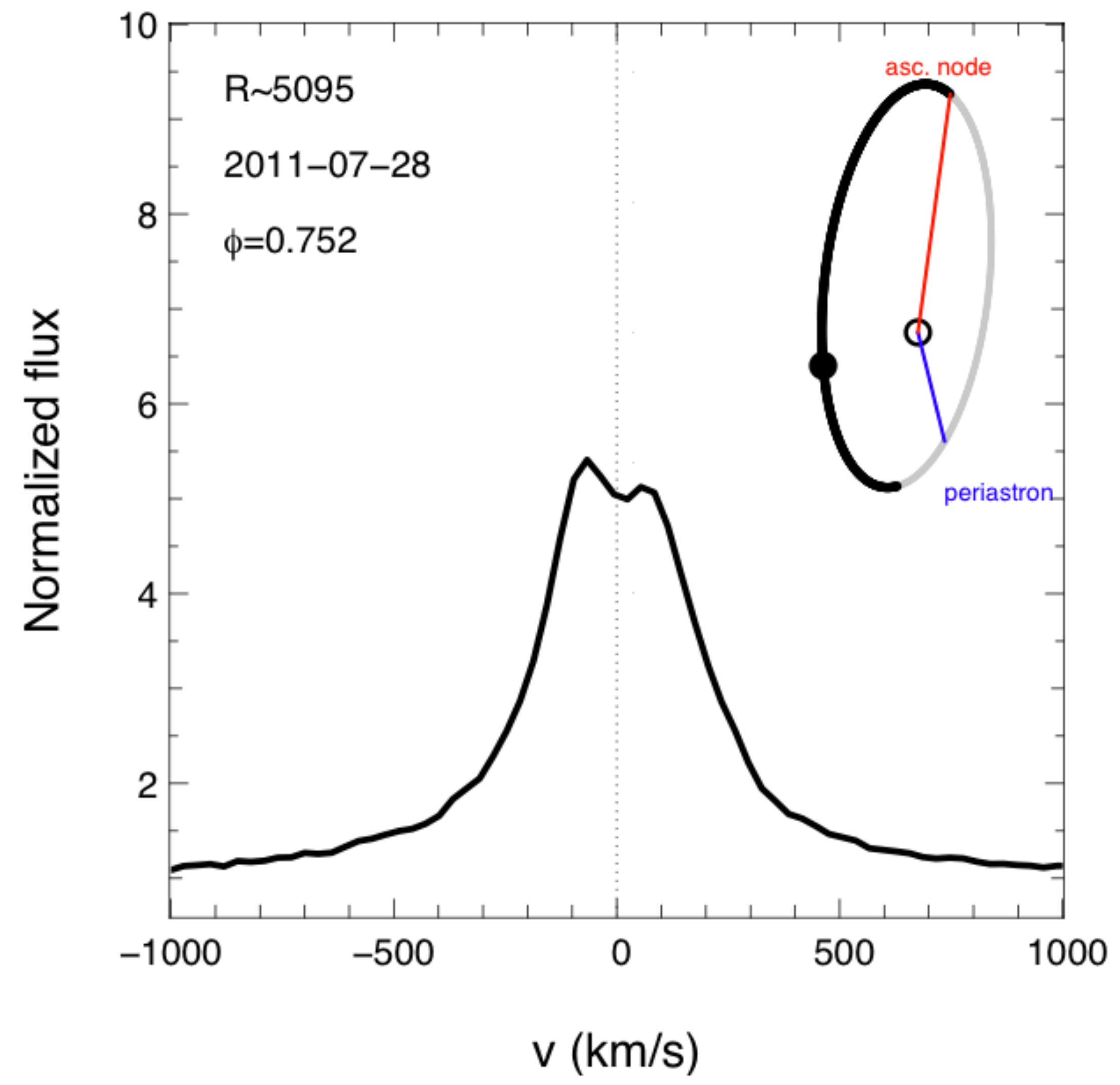} & 
\includegraphics[width=0.325\textwidth]{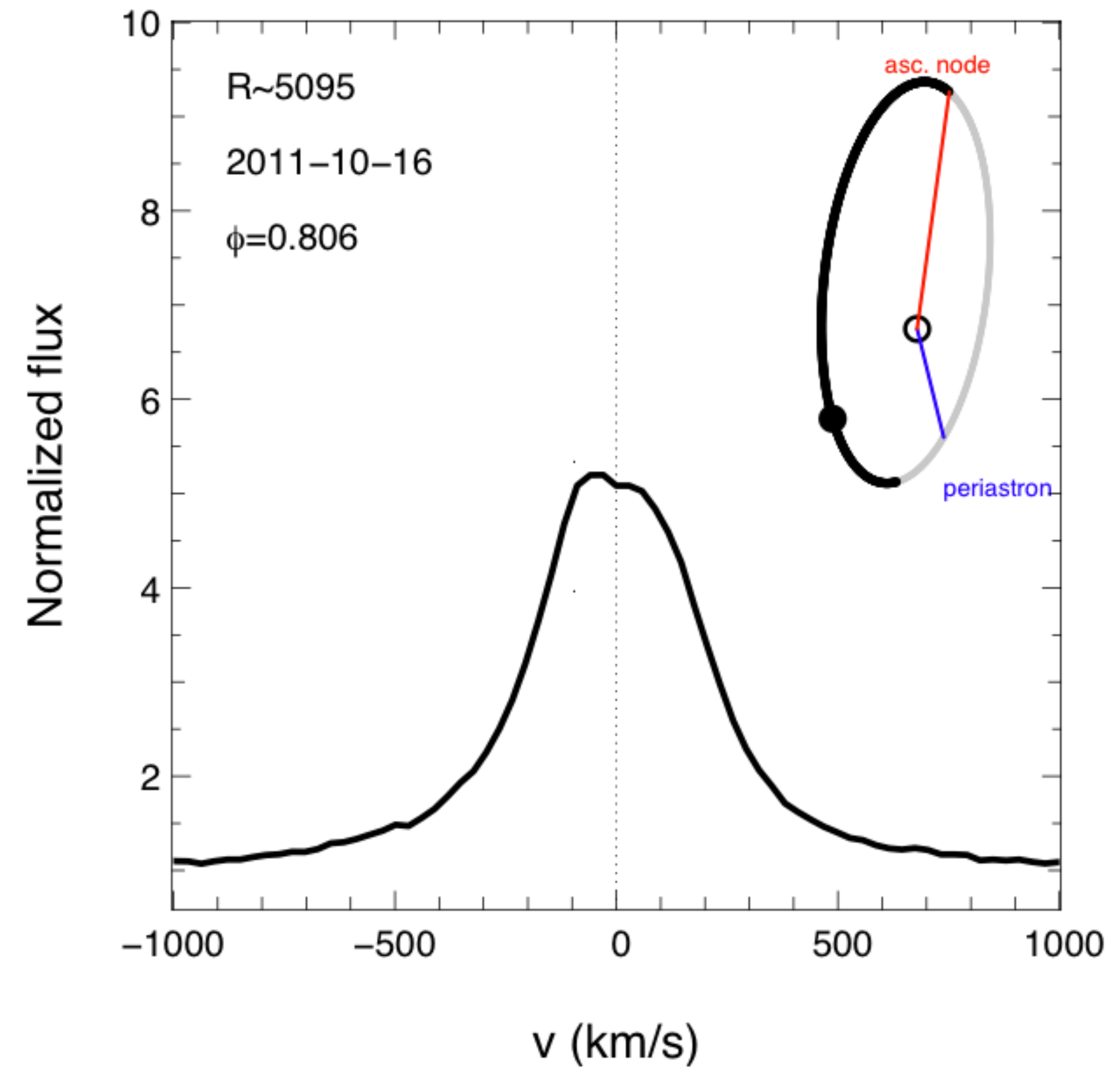} & 
\includegraphics[width=0.322\textwidth]{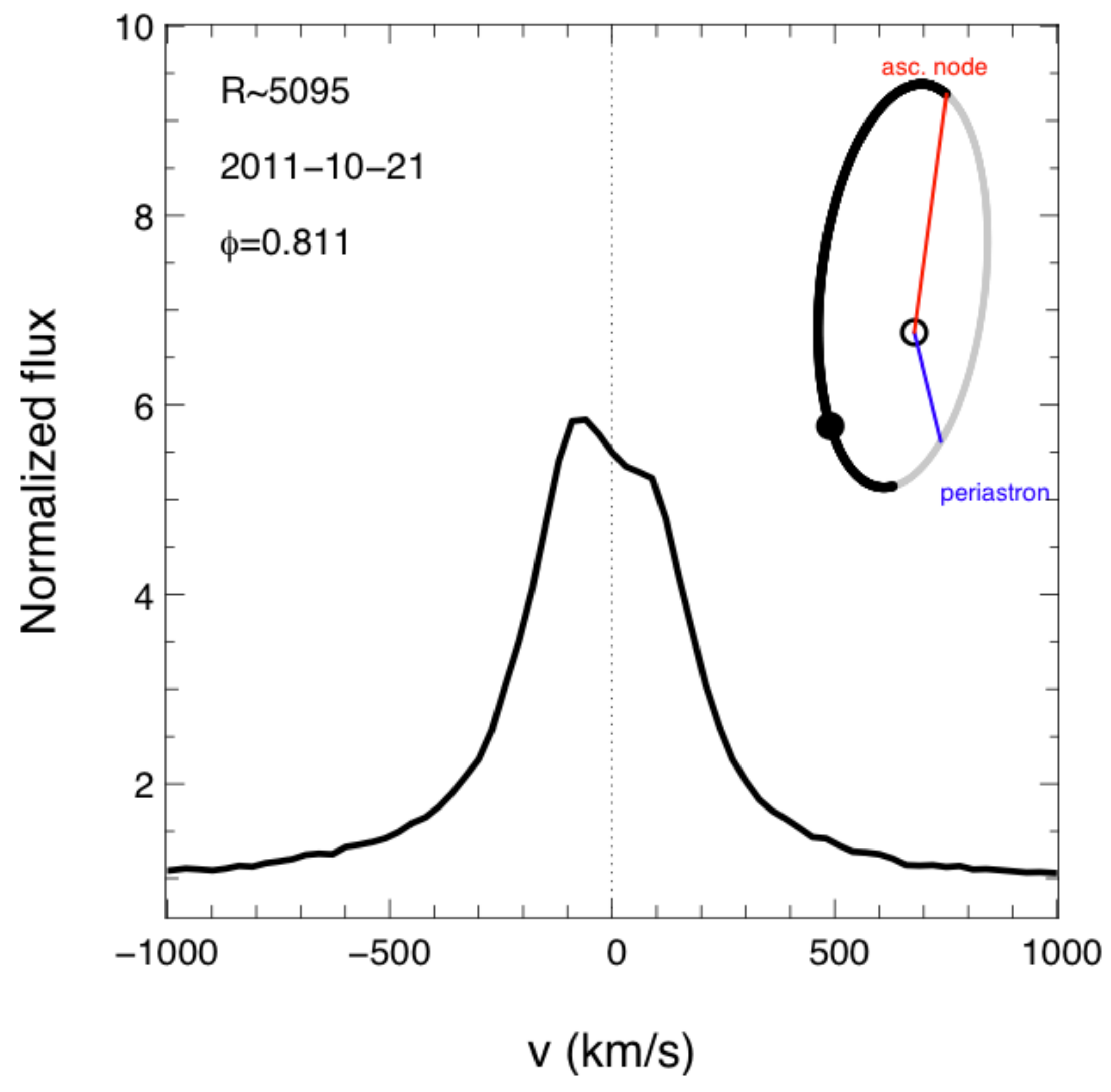}\\
\end{tabular}
\caption{Variation of the line profile measured with VEGA at R$\sim$5000, with the orbital phase. In each panel, the astrometric orbit is shown, with a full black circle indicating the corresponding orbital phase. The thick part is oriented towards the observer.}  
  \label{fig:spectreMR}
\end{figure*}

\begin{center}
\begin{table}
\caption{\label{SNR} Detection level of the V and  $\Delta\phi$ change in H$\alpha$. $\sigma_{\rm{cont}}$ and $\sigma_{H\alpha}$ are the errors in the continuum and in the line, respectively (computed as in Sect.~\ref{sect:dataproc}).}
\begin{tabular}{cccc}
  \hline
  \hline
  Date & $\phi$ & Detection V & Detection $\Delta\phi$ \\
  \hline
  \hline
 2008-07-29   & -0.012& 2.0 $\sigma_{\rm{cont}}$ / 3.2 $\sigma_{H\alpha}$ & 1.8 $\sigma_{\rm{cont}}$ \\
 2008-08-07   & -0.006 & 1.0 $\sigma_{\rm{cont}}$ / 1.7 $\sigma_{H\alpha}$ & 1.5 $\sigma_{\rm{cont}}$\\
 2009-08-26   & 0.262 & - / 1.0 $\sigma_{H\alpha}$ & - \\
 2010-08-01   & 0.499 & - / -  & 3.4 $\sigma_{\rm{cont}}$\\
 2011-08-28   & 0.773 & 1.5 $\sigma_{\rm{cont}}$ / 2.1 $\sigma_{H\alpha}$ & - \\
 2011-10-16   & 0.806 & 1.0 $\sigma_{\rm{cont}}$ / 1.8 $\sigma_{H\alpha}$ & 4.6 $\sigma_{\rm{cont}}$\\
  \hline
\end{tabular}
\end{table}
\end{center}

\section{Orbital parameters}
\label{sec:orb}

Several determinations of the binary orbit exist in the literature. However, the SB2 radial velocities \citep{alecian08} and resolved astrometric observations \citep{monnier06} have never been fitted conjointly. Consequently, we revisit these works with the goal of determining the dynamical distance of the system and the individual masses of the components. We combined the available datasets following the formalism detailed in \citet{lebouquin13} and performed a Levenberg-Marquardt least-square fit of the data. We are confident in our orbital determination, as the convergence toward a single and deep $\chi^2$ minimum is fast and robust for a wide range of initial guesses. The error bars on the parameters were obtained by bootstrapping. The best-fit parameters are provided in Table~\ref{tab:bestorbit}. We note that the dynamical distance is out of the confidence interval of the new reduction of the Hipparcos data ($520\pm150$\,pc, Van Leeuwen 2007) and the individual masses are smaller by a factor of 2 than the ones derived in \citet{alecian08}.

%\\\textbf{TBC: Should I mention something like that : Stellar parameters (temperature, radius) are very hard to guess. Consequently, the parameters in the literature are poorly reliable.}

\begin{table}[t]
  \caption{Best fit orbital elements and related physical parameters}
  \centering
  \begin{tabular*}{0.95\columnwidth}{@{\excs}lll}
    \hline\hline\noalign{\smallskip}
    Parameter & Value  & Unit\\
    \noalign{\smallskip}\hline\noalign{\smallskip}
    $t_\mathrm{p}$ &  $48962 \pm 55$ & {\tiny JD-2400000}\\
    $P$ &  $1433 \pm 17$  & days \\
    \Aap &  $15.9 \pm 0.7\:$ & mas \\
    $e$ &  $0.30 \pm 0.02$ & \\
    $\Omega$ &  $-7.7 \pm 6$&$\deg$\\
    $\omega$ &  $-136 \pm 1$&$\deg$\\
    $i$ &  $66 \pm 7$ & $\deg$ \\
    \Kp &  $19.8\pm 0.5$ & $\mbox{km\,s}^{-1}$ \\
    \Ks &  $16.1\pm 0.9$ & $\mbox{km\,s}^{-1}$ \\
    $v_0$ & $-8.1 \pm 0.3$  & km\,s$^{-1}$\\
    \noalign{\smallskip}\hline\noalign{\smallskip}
    \Aph &  $5.0 \pm 0.5$ & AU \\
    $d$ &  $320 \pm 51$&pc \\
    \Mp/\Ms &  $1.23 \pm 0.08$& \\
    \Mp+\Ms &  $9.8 \pm 3.6$& $\Msun$\\
    \Mp &  $5.37 \pm 1.9$&$\Msun$ \\
    \Ms &  $4.4 \pm 1.7$&$\Msun$ \\
    \noalign{\smallskip}\hline
  \end{tabular*}
  \label{tab:bestorbit}
\end{table}

%I used the CESAM evolutionary modes to obtain the evolutionary tracks expected for stars of these masses, that is Teff and R versus age. I used Kuruck models to build synthetic SEDs. I found the that the UV and visible flux (obtained via VOspec) can be reproduced by a system in late stage of PMS, assuming stars are coeval. But this is largely uncertain because of the unknown extinction.
%
%At best fit, the stars have parameters: $\Teffa=17500\,$K,  $\Ra=2.7\,\Rsun$, $\Teffb=14500\,$K and $\Rb=2.8\,\Rsun$ (which correspond to $\LogLa=2.77$, $\LogLb=2.47$). With those parameters, the flux ratio predicted in the V-band is $0.7$.
%
%\begin{itemize}
%\item Stellar masses are smaller than the one from Alecian.
%\item Distance is in the range $300-400\,$pc, most probably around $330\,$pc.
%\item Stellar parameters (temperature, radius) are very hard to guess. Consequently I believe the parameters in the literature are poorly reliable.
%\item The stellar parameters from Alecian are definitely not compatible with the evolutionary tracks of the stars.
%\end{itemize}

\section{Spectro-interferometry }
\label{sec:interfero}

\subsection{Spectroscopy}
Figure~\ref{fig:spectre} presents 14 different \Ha spectra that show a drastic change of intensity with time as well as slight changes in the full width at half maximum (FWHM) of the line. With these measurements,  we confirm that the equivalent width (EW) of the line (Fig.~\ref{fig:EW}) varies with the orbital phase, and reaches its maximum close to the periastron. In the following, we refer to 'active phase or state' when the system is close to the periastron,  otherwise to 'quiescent phase or state'.  In the quiescent state, we measure an EW of $\sim$35-40~{\AA}, about twice as much as in the active phase which is in agreement with previous studies \citep{miro98,alecian08}.
For the sake of clarity, Fig.~\ref{fig:spectreMR} shows six of the spectra obtained at medium spectral resolution, with a schematic of the system at the corresponding orbital epoch. The line is double peaked in the quiescent state with peak-velocities of $\sim$75-80~km/s, FWHM$\sim$450~km/s and broad wings at very high velocities ($\sim$800~km/s), as determined through Gaussian fitting of the profiles. The profile is almost symmetric close to the ascending node ($\phi\sim0.5$), and slightly asymmetric as the system gets closer to the periastron ($\phi\sim$0.8), with more blue-shifted emission. At periastron ($\phi\sim$0), the line is single peaked, and shows a slight asymmetry with more redshifted emission.  There is no variability in the broad wings. 
These results support the idea that the binarity is at the origin of the line profile and intensity variations, as expected and observed in other close binaries.

\subsection{Interferometry}
\label{sec:Viscont}
\begin{figure*}[!h] \centering
\includegraphics[width=1\textwidth]{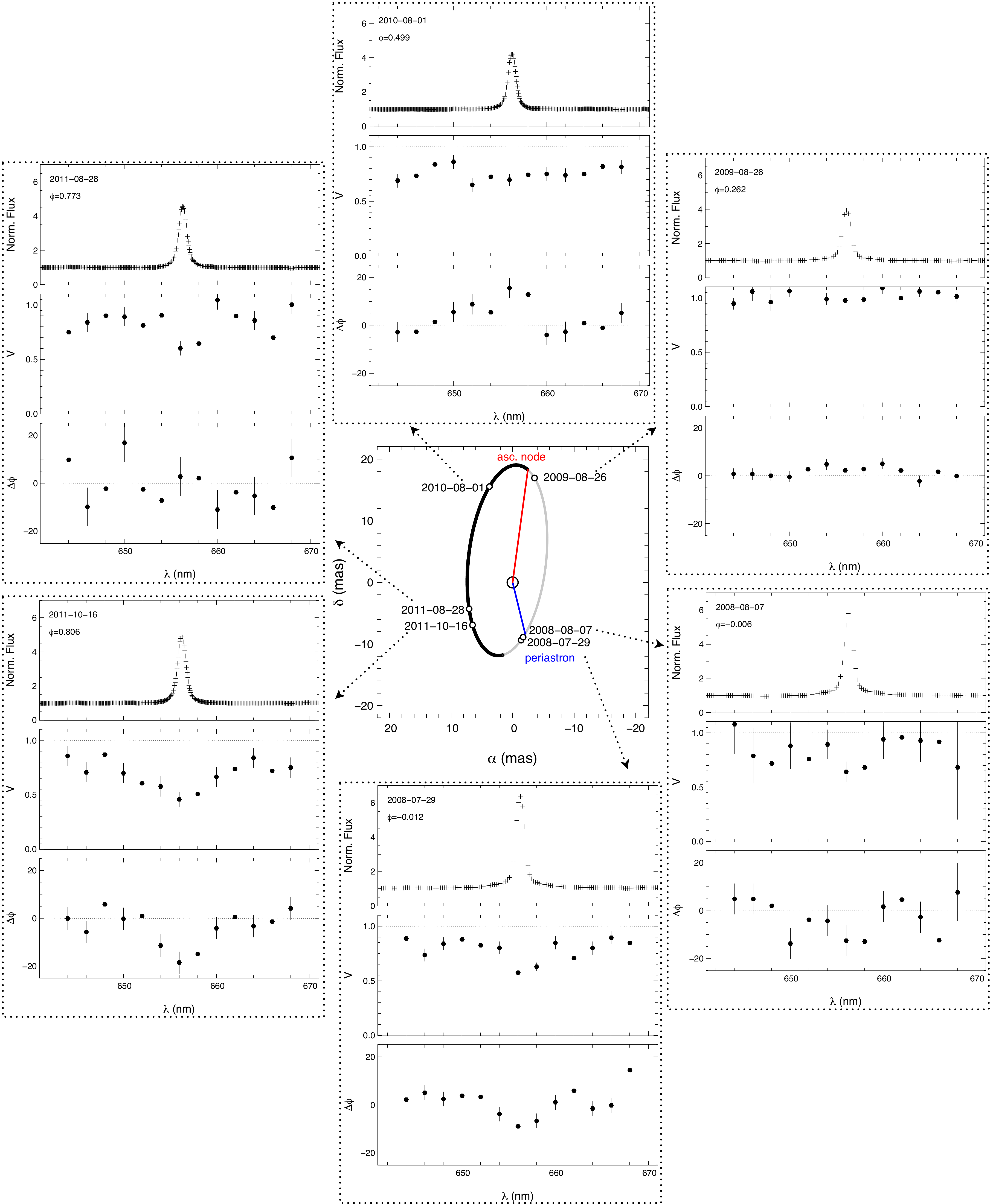}
\caption{VEGA spectro-interferometric measurements over the orbit. }
  \label{fig:VisvsOrbit}
\end{figure*}

Figure~\ref{fig:VisvsOrbit} presents the  visibilities and differential phases, as well as the corresponding spectra at a resolution of 1600.\\

\textit{Change in the spectral shape of the visibility.} By comparing the interferometric quantities in the continuum to the ones  in the spectral channel centred on H$_{\alpha}$, we  detect a marginal visibility drop in the line during the active states ($\phi\sim$0 and 0.8; Table~\ref{SNR}). This means that at least at these dates, the bulk of the \Ha emission is spatially resolved and extended. 

\textit{Change in the spectral shape of the  differential phases.} We find that within large error bars, the differential phase in the line is consistent with zero for all measurements, except for $\phi\sim$0.5 and 0.8. The differential phase signals can be translated into a photocenter displacement in the plane of the sky along the direction of the projected baseline. For the S1S2 baseline used for the observations, a positive phase corresponds to a photocenter displacement towards the south, and a negative phase corresponds  to the displacement  towards the north, as detailed in \citet{mourard12}. Therefore, the measured non-zero differential phases trace for all epochs a photocenter displacement in the direction opposite to the secondary.  As the measured differential phases are low ($\Delta\phi \leq 20$\dg), the photocenter of the \Ha emission is rather close to the continuum/binary photocenter.   Detection levels are defined using the errors in the continuum and in the line ($\sigma_{\rm{cont}}$ and $\sigma_{H\alpha}$, respectively, computed as in Sect.~\ref{sect:dataproc}) and are reported in Table~\ref{SNR}.

\textit{Change in the absolute level of the visibility.}  We notice that the level of the continuum visibility varies with the orbital phase, even if the measurements were obtained at similar angular resolutions ($\sim$2.5~mas). This behavior is expected since the continuum emission is due to the binary that is spatially resolved by the interferometer. Therefore, any change in binary separation, position angle and/or flux ratio strongly affects the visibilities. Close to the ascending node (2009 dataset), we measure a continuum visibility close to 1, corresponding to an unresolved emission along the baseline position angle, while in the active state, the continuum visibility can be as low as $\sim$0.76.

\begin{table*}[t] 
\centering 
\caption{Orbital phase ($\phi$), angular separation ($\Delta \alpha $,  $\Delta \delta $), and position angle (PA$_{\rm{bin}}$) of the SB at the date of the interferometric observations. PA$_{\rm{B}}$ is the baseline position angle, FR the binary flux ratio in the continuum, $\Theta_{\rm{H_{\alpha}}}$ and $\rho_{\rm{H_{\alpha}}}$  the \Ha FWHM and displacements resulting from the best fit of a face-on Gaussian model, in mas and AU at 320~pc. The displacements are counted positive towards the secondary.}
\label{tab:vis}
\begin{tabular}{ccccccccccc}
\hline
Date & $\phi$ & $\Delta \alpha $  & $\Delta \delta $ & PA$_{\rm{bin}}$ &  PA$_{\rm{B}}$   & FR & $\Theta_{\rm{H_{\alpha}}}$ & $\Theta_{\rm{H_{\alpha}}}$ & $\rho_{\rm{H_{\alpha}}}$&$\rho_{\rm{H_{\alpha}}}$  \\
& & (mas) & (mas) & (\dg) &  (\dg) & &  (mas) & (AU) & (mas) & (AU)\\
\hline
2008-07-29  & -0.012 & -1.372&  -9.387 &-171.7 & -31.0 & 0.10$^{+0.03}_{-0.04}$ & 1.91$\pm$0.13 &0.61$\pm$0.04 & -0.10$\pm$0.03 &  -0.03$\pm$0.01\\
2008-08-07  & -0.006 &  -1.712&  -8.891 &-169.1 & -28.0   & 0.07$^{+0.05}_{-0.06}$ & 1.65$\pm$0.33 & 0.53$\pm$0.11 &-0.19$\pm$0.03 & -0.06$\pm$0.03\\
2009-08-26  & 0.262 & -3.558& 16.961 & -11.8 & -31.0   &0.01$^{+0.05}_{-0.01}$  &0.67$\pm$0.20 &  0.22$\pm$0.06 &-0.03$\pm$0.04 & -0.01$\pm$0.01\\
2010-08-01  & 0.499 &  3.766&15.567 &  13.6& 0.9  & 0.29$^{+0.18}_{-0.11}$ &   1.41$\pm$0.17 & 0.45$\pm$0.05 & 0.0$\pm$0.10&0.0$\pm$0.03\\
2011-08-28  & 0.773 & 7.066 & -4.305 &121.3   & -44.0  &0.07$^{+0.08}_{-0.06}$ &  1.95$\pm$0.20 & 0.62$\pm$0.06 & 0.4$\pm$0.04& 0.0$\pm$0.04\\
2011-10-16  & 0.806 &  6.510&-6.927 & 136.8 & -36.0  &  0.99$^{+0.01}_{-0.98}$ & 1.85$\pm$0.20 & 0.59$\pm$0.06 & -0.32$\pm$0.30&-0.10$\pm$0.10 \\
%2011-10-21  &  0.860 &  3.237& -10.629 &163.1 & 92.0  & / & / & / \\
\hline
\end{tabular}
\end{table*}

%    
%    
%> Rho_mas
%[-0.105528,-0.165829,0,-0.120603,0,-0.316583]
%> Rho_au
%[-0.0337688,-0.0530653,0,-0.038593,0,-0.101307]
%> 
%
%    
%"Chi2r Visibilite= [1.06566,0.0678234,0.0184719,0.538924,0.194033,0.462456]"
%"Chi2r Phase= [0.318584,0.00464858,3.58359,0.228782,0.311073,0.284893]"
%"Chi2r TOTAL= [1.11226,0.0679825,3.58364,0.585475,0.366626,0.543167]"
%"Chi2r Visibilite L+C= [0.840271,0.16834,0.579028,1.10589,0.873079,1.48251]"
%"Chi2r Phase= [2.54418,0.843549,1.29541,0.881167,1.09253,0.86852]"
%"Chi2r TOTAL= [2.67935,0.860183,1.41893,1.41401,1.39853,1.71818]"
    
\section{Results}
\label{sec:res}

The goal of the analysis is to derive the basic properties of the \Ha emitting region. This requires a proper subtraction of the continuum emission that is only due to the binary stars. To do so, we first fit a binary model to the continuum visibilities, with fixed orbital parameters (Table~\ref{tab:bestorbit}), in order to determine the flux ratio (FR) between the two stars. This flux ratio will set our model in the continuum for both the visibilities and differential phases. 

\subsection{Binary flux ratio}
In the spectral range of our observations, we assume that the only emitters in the continuum are the two stellar components. Their relative flux ratios, separations, and position angles have a strong impact on the visibility in the continuum. Assuming that the relative stellar fluxes vary over the orbit, we can use our continuum visibilities to retrieve the flux ratio at different epochs, assuming that the individual stars are unresolved, and using a simple analytical formula for the binary visibility. 

We first compute the position angle and separation for each epoch, using the orbital parameters determined in Tab.~\ref{tab:bestorbit}. We then compute a large grid of $\sim$1000 binary models, and minimize a $\chi^2$. Individual fits to each data set provide flux ratios from 0.07$^{+0.05}_{-0.06}$ to 0.29$^{+0.18}_{-0.11}$ ($\chi^2$=1.5 at most). This suggests that the secondary is the dimmer object in the continuum. At the ascending node $\phi\sim$0.26 where V$\sim$1 and the binary is unresolved, all flux ratio values below 0.06 provide an equally good fit ($\chi^2$=1.1), while for $\phi\sim$0.81 all possible flux ratio values surprisingly provide a bad fit ($\chi^2$=3.3). The second dataset shows a large intrinsic scatter in the continuum  (rms$\sim$10\%), and seems to suffer from calibration problems that prevent us from determining a good absolute value for the visibility. Because we cannot determine a flux ratio for this measurement,  we scale the average continuum visibilities, for this dataset only, to the binary model predictions with FR=0.10, which corresponds to the best-fit model assuming a constant flux ratio over the entire orbit ($\chi^2$=3.8). The scaling factor is $\sim$25\%, as shown in Fig.~\ref{fig:appendix}, together with the best fit of the binary model for each dataset.  Values of flux ratios are given in Table~\ref{tab:vis}. Because of the large error bars, we cannot determine whether the change in FR with the orbital period is significant and expect that further high SNR observations would answer this question.

%Since the large error bars does not allow us to consider whether the changes in flux ratio between the various epochs are significant or not, 

\subsection{Size and photocenter of the H$_\alpha$ emission}
To model the \Ha circumstellar emission, we consider the contribution of the binary to the measured visibilities and differential phases  using the complex visibilities. Assuming that the binary is at the photocenter of the continuum emission (i.e., $\Delta\phi$=0 in the continuum), we find that for each spectral channel $k$ 

\begin{equation}
\rm{V}_{k}  e^{i\Delta\phi_{k}} = \frac{F_{cont,k}V_{cont,k} + \rm{F_{H_{\alpha},k}} \rm{V_{H_{\alpha},k}} e^{i\Delta\phi_{H_{\alpha},k}} }{ F_{cont,k} + \rm{F_{H_{\alpha},k}}},
\label{eq1}
\end{equation}

\noindent where the indices cont and $H_{\alpha}$ refer to the continuum (binary) and circumstellar emission, $\Delta\phi_{k}$ is the measured 
differential phases in the spectral channel $k$, and $\Delta\phi_{H_{\alpha}}$ is the differential phase due to the \Ha emitting region. The real and imaginary parts of Eq.~\ref{eq1} lead to two equations that can be solved for $\rm{V_{H_{\alpha}}}$ and  $\Delta\phi_{H_{\alpha}}$ \citep[see, eg.,][]{weigelt07,eisner10}.  

\begin{figure}
 \centering
\includegraphics[width=0.5\textwidth]{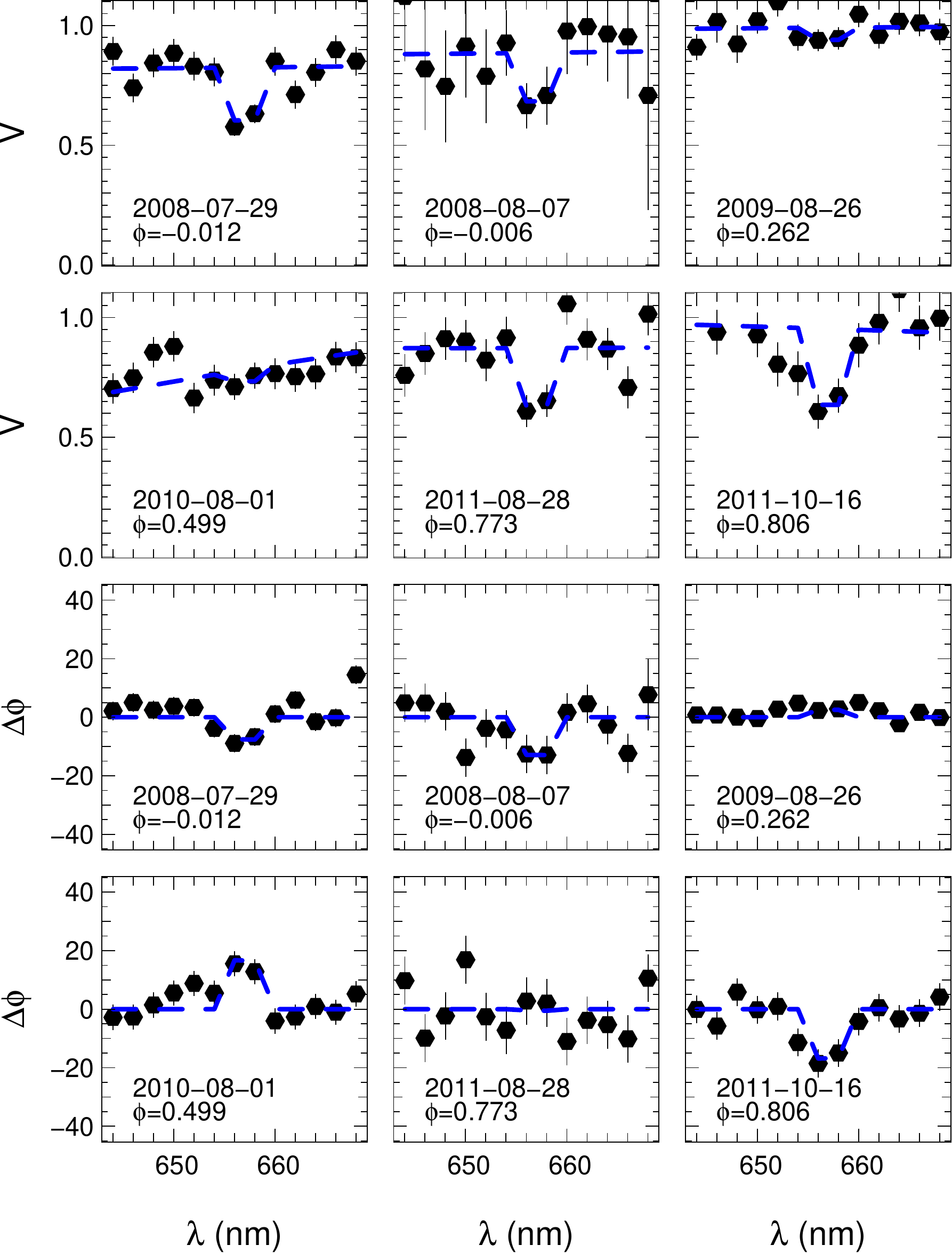} 
\caption{Best-fit continuum and \Ha line model (dashed line). The first two lines give the visibilities, the following two, the differential phases. We recall that the visibilities for $\phi$=0.806 have been scaled to the binary model with FR=0.10.}
\label{fig:bestfit}
\end{figure}

\begin{figure}[t] \centering
\includegraphics[width=0.4\textwidth]{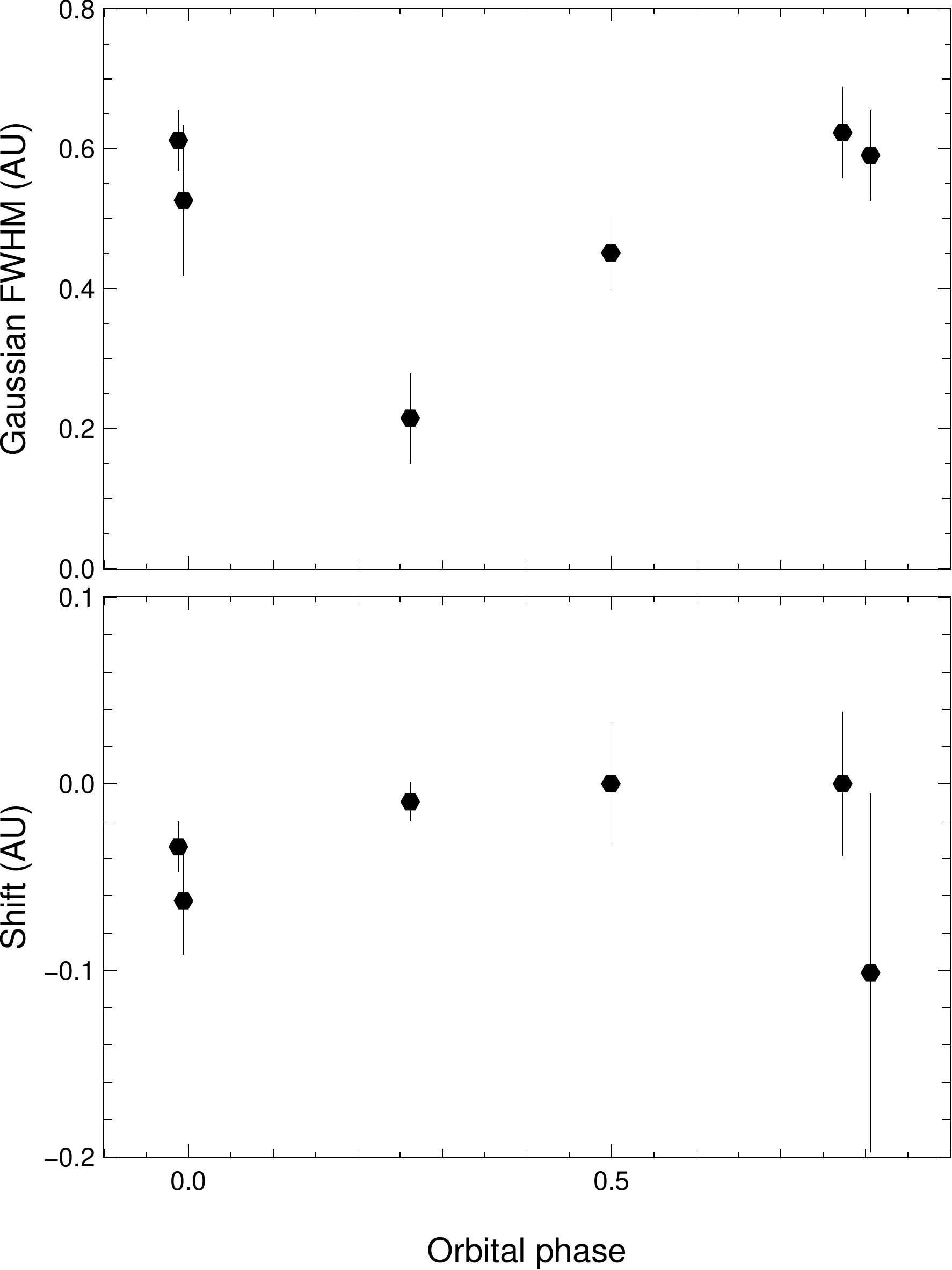}
\caption{Best-fit model parameters, Gaussian FWHM (top) and displacement (bottom), over the orbit.}
  \label{fig:G}
\end{figure}

With the best binary model, and the line to continuum ratio determined from the spectra after subtraction of the photospheric \Ha absorption, we fit our model for the \Ha emission to $\rm{V_{H_{\alpha}}}$ and  $\Delta\phi_{H_{\alpha}}$, in the two spectral channels that contain most or all of the line. Considering the low quality of our datasets, we limit ourselves to a simplistic analytical approach with a face-on Gaussian model, which FWHM (considered as a typical size) and location can vary. We restrict the Gaussian displacement along the binary axis only, since our measurements have been obtained along a single baseline orientation at a time, which prevents us from determining a full 2-D position. The reference is taken to be the primary and  displacements are counted positive towards the secondary. %We consider a face-on (i=PA=0\dg) model and a model coplanar with the circumbinary disk (i=54.5$^\circ$, PA=6.5$^\circ$ \citet{okamoto09}). 
We vary the Gaussian FWHM from 0 to 6 mas, and since the sign of the differential phases indicates a displacement in the  direction opposite to the secondary, we vary the Gaussian location from -4 to 0 mas, and minimize a $\chi^{2}$ to find the best model parameters.\\
\indent Figure~\ref{fig:bestfit} shows the best-fit to the \Ha visibilities and differential phases,  and Fig.~\ref{fig:G} presents the model parameters as they vary with the orbital phase (values are in Table~\ref{tab:vis}). Our model fits the observations well, with a reduced $\chi^{2}$ of 2.7 (at most) in the line+continuum (1.1 in the \Ha spectral channels only).  Conservative error bars have been obtained by considering the extreme values of the visibilities and FR within their error bars. \\
\indent We find that the spatial extent of the emission changes over the orbit. In the quiescent state, the Gaussian FWHM is 0.67$\pm$0.20 mas (i.e., $\sim$0.22~AU) and increases up to $\sim$1.95$\pm$0.20 mas (i.e., $\sim$0.62~AU), close to periastron.  The displacements towards the opposite direction than the secondary appear to be quasi constant along the orbit, with absolute values close to 0.10$\pm$0.03~mas (i.e., 0.03$\pm$0.01~AU). The errors on the displacement are dominated by the errors on the binary flux ratio, hence are very large for $\phi\sim$0.81.

%At the ascending node, the displacement estimate is lower (at the 2-$\sigma$ level) than at other orbital phases  (absolute values of 0.14$\pm$0.02~mas, i.e., 0.05$\pm$0.01~AU). This result is consistent with a smaller and more compact emitting region. The last measurement ($\phi$=0.86) appear to differ from the others, and in particular with the measurement obtained at a similar phase $\phi=0.83$. This is  probably due to the small difference in baseline orientations, as models with a slightly different morphology (e.g., non-zero inclination) provide consistent results.

Because our  measurements only probe one direction at a time, we are not able to determine an inclination and a position angle for the \Ha emitting region, and leave this for a detailed radiative transfer modeling combined with additional simultaneous multi-baseline interferometric observations. Thus, we would like to stress that the obtained parameter values are model-dependent and different geometries (e.g., ring, uniform disk) lead to slightly larger values of typical size, as expected with such simple prescriptions. In addition, the values of the displacement strongly depend on the binary flux ratio that set the phase values in the continuum. However, the variations of the \Ha characteristic size along the orbit and the negative photocenter displacements still hold.\\

%we can derive an estimate of the location of the dust sublimation radius of the circumprimary disk to be between 3.7 and 8.4~AU \citep[e.g.,][]{dullemond10}, using~:}
%\begin{equation}
%\label{sub}
%R_{\rm{sub}} \sim \sqrt{\frac{L_{\rm{*}}}{4 \pi \sigma T_{\rm{sub}}^{4}}},
%\end{equation}
%
%with T$_{\rm{sub}}$=1500~K, the typical dust sublimation temperature for silicates. 
%
%using Eq.~\ref{sub}, but comparable with the dust inner radius ($\sim$1.73~mas / 0.62~AU) measured by \citet{monnier06}. This apparent discrepancy can be explained with optically thick material inside the dust sublimation radius that efficiently shields the dust from the stellar light  \citep{monnier02}. 

\section{Discussion}
\label{sec:disc}

\subsection{Characteristic spatial extents in \Ha}

It is useful to compare the \Ha size estimates to typical radii in the close circumstellar regions probed by our measurements. If we consider the extreme values of luminosities found in the literature ($\sim$3000 and $\sim$15000~L$_{\odot}$ from \citet{alecian08,hernandez04}), we derive an estimate of the location of the dust sublimation radius of the circumprimary disk to be between 3.7 and 8.4~AU \citep[e.g.,][]{dullemond10}. These estimates are larger than the dust inner radius ($\sim$1.73~mas, i.e., 0.55~AU at 320~pc) measured by \citet{monnier06}. This apparent discrepancy can be explained with optically thick material inside the dust sublimation radius that efficiently shields the dust from the stellar light  \citep{monnier02}. 

Considering the vsin(i) estimate and stellar properties from \citet{alecian08}, we find that the corotation radius should be $\sim$0.30$\pm$0.15~AU. This estimate suffers from the large uncertainties on the stellar parameters inferred in the literature. Nonetheless, our Gaussian fit seems to indicate that the bulk of the \Ha emission as described by a Gaussian model is most likely located between the corotation radius and the dust sublimation radius, and is more extended than the typical regions involved in accretion processes onto the star. In the case of a low mass star, the truncation radius is set by the interaction of the stellar magnetic field and the gaseous disk in a region very close to the star. In a massive star, although this is still a matter of debate, the accretion is thought to occur through a boundary layer. 

Similar extents for Hydrogen line emission have been found in other Herbig~Be stars, in the Br$_{\gamma}$ line, and interpreted as originating in a stellar or disk wind \citep[e.g.,][]{kraus12,weigelt11,benisty10}, in contrast with the findings of \citet{eisner10} on a survey of lower mass young stars that are consistent with accretion.

It also seems unlikely that the bulk of the \Ha emission is due to a rotating disk, even in the quiescent state, where the \Ha line profile shows a double peak at low velocities ($\sim$75~km/s). If we consider a mass of $\sim$10~M$_{\odot}$, and the velocity field of a disk in Keplerian rotation ($v=\sqrt{G M_{*}/R}$/sin(i)), the \Ha double peak in the quiescent state could result in a rotating disk with a minimum outer radius of R$\sim$1.2~AU, a much larger value than our estimates (Table~\ref{tab:vis}). The disk region responsible for the broad \Ha wings would be located very close to or inside the stellar surface ($\sim$7.6~R$_{\odot}$).

\subsection{Origin of the \Ha burst}

The results given in the previous section can be interpreted in the context of mass ejection. The extended \Ha emission together with the increase of the \Ha line intensity may trace a period of enhanced mass loss in a strong wind emitted by the primary. Such an event could follow a period of enhanced mass accretion from the circumbinary disk to the primary disk, triggered at periastron as predicted by numerical models. The photocenter displacements in the direction opposite to the secondary indicate that if a wind is responsible for the \Ha emission, the emission should be enhanced along this direction. The wind therefore could  not be fully spherical, as the consequent emission would be symmetric. 

Figure~\ref{fig:spectre} shows that the \Ha broad line wings ($\sim$800~km/s) are very stable, and can be due to outflowing matter, as such velocities are often seen in wind tracers. The additional emission at periastron is produced only at very low projected velocities, probably coming from a plane close to perpendicular to our line of sight. The lack of P~Cygni absorption indicates that the wind is not self-absorbing along our line of sight. This would support a wind with a higher density towards low latitudes, close to the equatorial plane, in agreement with the model predictions of poorly collimated winds in massive stars \citep{vaidya11}.  With a non-zero inclination, an outflow can also produce the double peaked profile observed in quiescent state, as shown in \citet{stee94} for Be stars, and more recently in \citet{weigelt11} for a Herbig~Be star of similar mass. However, complex line opacity effects can come into play, as suggested by  \citet{elitzur12} who showed that it is in practice impossible to disentangle effects of kinematics and line opacity effects in double peaked lines and that proper radiative transfer is needed. In addition, we note that spectra obtained at high spectral resolution \citep{alecian08} show that the line profile is extremely complex with variable features, which seem to indicate a combination of various variable mechanisms.

The additional emission at periastron could, in principle, come from either the primary, the secondary, or a combination of both. However, the low differential phases indicate that the photocenter of the \Ha emitting region is  close to the continuum photocenter, which is located close to the primary (since the binary best-fit flux ratio is 0.10). The displacement values found by our simple analytical model fitting therefore rule out a scenario in which all of the additional \Ha emission at periastron comes from the secondary. It seems therefore very unlikely that accretion onto the secondary or wind-enhancement on the secondary is the key-mechanism at periastron or that both objects are equally responsible for the increase. On the contrary, because the displacements are in the opposite direction to the secondary, the burst of activity at periastron seems to be connected to the primary. Although the variation of photocenter with the orbital phase is marginal, a smaller displacement at the ascending node is consistent with a smaller emitting region and a lower line intensity. An almost-constant displacement at all other phases is also in agreement with the derivation of a similar extent at these dates. This is supported by recent numerical simulations of accretion flows in close binaries done at high resolution \citep{devalborro11}. These models suggested that  accretion from the circumbinary disk onto the primary is favored at periastron. In that scenario, it is expected that most or all of the additional H$\alpha$ emission in the active state comes from the primary star, as our data seem to indicate. We note, however, that with our single-baseline measurements, we are not able to disentangle more complex scenarios.

Whatever the mechanism favored to enhance the \Ha emission at periastron, additional mass has to be fed into the innermost regions at these active stages. If mass is transferred through accretion streamers from the circumbinary disk to the primary disk, there should be a consequent time delay to allow matter to flow through the primary disk and reach the innermost AU. The viscous time is proportional to $\sim$R/ ($\alpha$ c$_{\rm{s}}$ H/R), with R the radius, H the disk scale height, and $\alpha$ the turbulence parameter (proportional to the viscosity; \citet{shakura73}). Considering a maximum outer disk radius for the circumprimary disk of 2~AU (because the binary separation at periastron is $\sim$4~AU, \citep{monnier06}), $\alpha$=0.01, and H/R=0.05, the viscous time needed for the mass flow to propagate is of the order of a few ten thousand years, much larger than the binary period. An alternative scenario is that the secondary induces gravitational instabilities in the primary disk as it approaches. In fact, tidal effects could induce strong waves all over the disk spatial range and produce instabilities, which would in turn enhance the mass accretion in the inner AU and result in a higher mass loss. The exact physical process that leads to the \Ha increase still remains to be understood.

The comparison of Fig.~\ref{fig:EW} and Fig.~\ref{fig:G} seems to suggest that the variations in \Ha EW and sizes are slightly shifted in time, as the size appears to increase faster than the absolute EW. This delay is difficult to interpret as we would expect a larger emitting area to produce a stronger \Ha emission, but considering the large error bars on our size estimate and the sparse time sampling of the orbit, we cannot determine whether it is a real feature or not.

%\textbf{Time it takes to go from quiescent to active state (and its possible  relation with stellar distances) both the EW and the size : }active to quiescent in both EW and size is from $\phi$=0.1 to 0.3 but slower increase of the EW from quiescent to active state ($\phi$=0.3-1) than in size (size lower, but consistent within large error bars at $\phi$=0.6). However, we have sparse measurements in EW and sizes, so not sure that's very relevant. But our data seem to indicate a time delay. This is difficult to explain, and more observations would be needed to quantify this.  \\
%\textbf{You could possibly compare those timescales that with e.g simulations of Gunther and  Kley (2002) that provide accretion rates evolution as a function of  orbital period (fig 11. and 12.). }\\

%\textbf{Do you see a possible time delay between periastron passage and maximum activity  (as suggested by figure 2) ? }: Yes this is also predicted in de Val Borro - right ? But we can't say for sure, because we don't have enough time sampling to say whether it's at phase=0 or phase = 0.1 that we see an increase. Same Alecian, not a continuous monitoring.   \\
%\textbf{- If you want to compute the viscous timescales rigorously you should 
%use the maximum size disk as determined by the stellar roche lobe 
%size. Since you have the mass ratio you can compute the position of 
%the lagrangian points. }\\

%\textbf{Discussion about accretion ? What if we are witnessing the \Ha emission in the boundary layer? }

\section{Summary and conclusion}
\label{sec:conc}
In this paper, we report the first spatially and spectrally resolved observations of the \Ha line emitting region in a close binary system using the VEGA spectrometer on the CHARA array.  As previous authors have already pointed out \citep{pogodin04,alecian08}, we find that \hd~shows a variation of the \Ha EW with the orbital phase. We use a simple Gaussian model located along the binary axis to fit the \Ha visibilities and differential phases, and find marginal evidence for a change in spatial extent during the orbit that is minimum in quiescent state (close to the ascending node). The \Ha Gaussian photocenter is at all epochs located in the direction opposite to the secondary, supporting a dominant emission by the primary. 

We find typical \Ha extents  within the sublimation radius. These  \Ha spatial extents do not support an accretion origin for the bulk of the line emission, as it would involve  smaller scales. Instead, we interpret these results in the context of an enhanced mass-loss event, triggered by the gravitational influence of the secondary. Such a scenario would suggest a strong connection between accretion and ejection in this Herbig~Be star, as already noted in other massive objects \citep[e.g.,][]{ellerbroek11,benisty10}.

A dominant contribution to the \Ha line by the primary is supported by our results and by the analysis of \citet{alecian08} and \citet{monnier06}. However, although the stellar parameters are still highly uncertain, it is puzzling that only the primary has a circumstellar disk which accretes from the circumbinary disk. In particular, the kilo-Gauss magnetic field detected on the secondary should induce the formation of a gaseous disk which should in turn emit H$_{\alpha}$. The circumstellar environments of the two objects must either have evolved differently, or the secondary could itself be an unknown spectroscopic binary of lower mass which would more easily explain the strong magnetic field, an uncommon feature in Herbig~Be stars.

We find that the binary flux ratio in the continuum is $\sim$0.10 (from $\sim$0.07 to $\sim$0.29, Tab.~\ref{tab:vis}). Because of large error bars on the visibilities, we are not able to retrieve a precise flux ratio at each orbital phase and to determine whether the change in \Ha EW is correlated with some change in the amount of reprocessed and scattered light as observed in other young close binaries \citep{vanboekel10}. We notice a discrepancy between our flux ratio estimates in the continuum and the estimate of \citet{alecian08}, based on an analysis of spectral lines ($\sim$0.67). These estimates are difficult to reconcile, unless one invokes variable extinction on one of the two objects, or if one of the two (most likely the magnetic secondary) is already chemically peculiar. 
Besides, the large uncertainties on the individual masses, the distance, and age of the system prevents the nature of the individual objects from being assessed precisely. As noted by \citet{alecian08}, photometric observations of each component separately are necessary. 
In addition, a regular and full coverage of the orbit is required to answer a number of open questions, such as a possible time delay between periastron passage and maximum activity.   More detailed modeling, as well as an intensive high SNR spectro-interferometric campaign with additional baselines at various orientations are also expected to properly constrain the complex interplay between accretion and ejection in such massive close binary systems. \\

\appendix
\section{Appendix}

\begin{figure}[h] \centering
\includegraphics[width=0.4\textwidth]{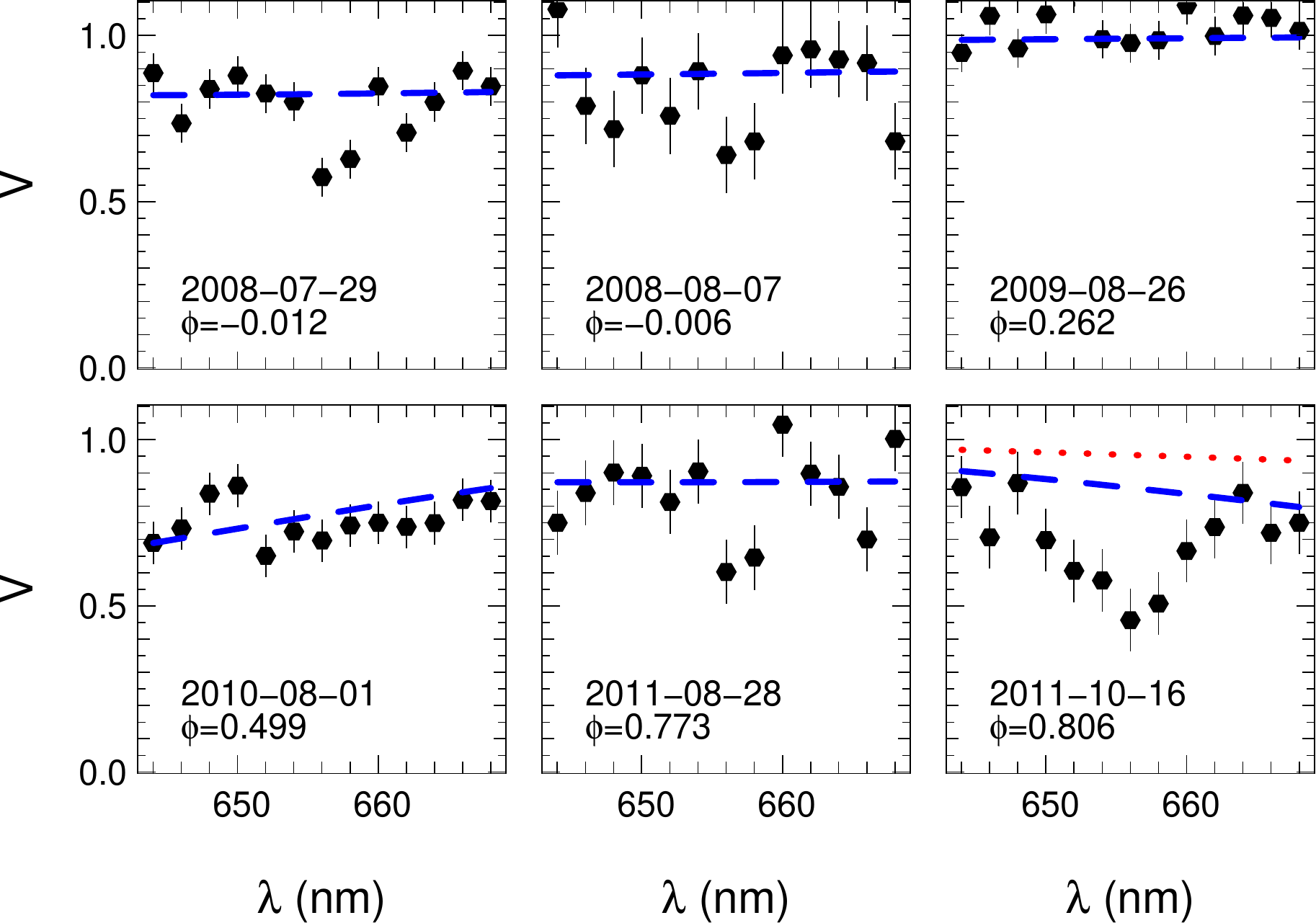}
\caption{Best-fit binary models in the continuum (dashed blue line). The red dotted line in the lower right panel is the prediction of a binary model with flux ratio of FR=0.10 to which the continuum visibilities measured for $\phi$=0.806 were re-scaled.}
  \label{fig:appendix}
\end{figure}

\begin{acknowledgements}
We acknowledge the anonymous referee for the constructive report that helped improve the manuscript.  We thank E.~Alecian, J.~Bouvier,  P.~Garcia, A. M\'erand, J.L.~Monin and J. Monnier for fruitful discussions. We acknowledge all the great observers at the CHARA Array who helped to gather this large dataset. The CHARA Array is operated with support from the National Science
Foundation through Grant AST-0908253, the W. M. Keck Foundation, the
NASA Exoplanet Science Institute, and from Georgia State University. GHRAL acknowledges CAPES for the support. This research has made use of the Jean-Marie Mariotti Center \texttt{SearchCal} service \footnote{Available at http://www.jmmc.fr/searchcal}
co-developped by FIZEAU and IPAG, and of CDS Astronomical Databases SIMBAD and VIZIER \footnote{Available at http://cdsweb.u-strasbg.fr/}.
\end{acknowledgements}

\bibliographystyle{aa}
\bibliography{mwc361}

%\begin{thebibliography}{} %\bibitem[Alecian et al. 2008]{Alecian} Alecian, E., Catala, C., Wade, G.A., %   et al., 2008, MNRAS, 385, 391 % %\bibitem[Brummelaar et al. 2005]{chara} ten Brummelaar, T. et %al. 2005, ApJ 628, 453 % %\bibitem[Fuente et al. 1998]{fuente} Fuente, A., Martin-Pintado, J., Rodriguez-Franco, A., et
%al., 1998, A\&A, 339, 575 % %\bibitem[Li et al. 1994]{LI} Li, W., Evans, N. J., Harvey, P. M., %\& Colome, C. 1994, ApJ, 433, 199 % %\bibitem[Millan-Gabet et al.io2001]{MG01} Millan-Gabet, R., Schloerb, %F.P., Traub, W.A. 2001, ApJ, 546, 358 % %\bibitem[Miroshnichenko et al. (1998)]{Miro98} Miroshnichenko, A.S., Fremat,
%Y., Houziaux, L. et al. 2006, PASP, 110, 883 % %\bibitem[Monnier et al. 2006]{Monnier} Monnier, J.D., Berger, %J.P., Millan-Gabet, R.G. et al. 2006, ApJ, 647,444
% %\bibitem[Mourard et al. 2009]{vega} Mourard, D., Clausse, J.M., %Marcotto, A., et al. 2009, A\&A, 508, 1073 % %\bibitem[Pogodin et al. 2004]{pogo} Pogodin,
% M.A., Miroshnichenko, A.S., %Tarasov, A.E., et al. 2004, A\&A, 417, 715 % %\end{thebibliography}{}

\end{document}